\documentclass[journal]{IEEEtran}
\usepackage{algorithm}
\usepackage{algorithmicx}
\usepackage{graphicx}          
\usepackage{mathrsfs}         
\usepackage{amsmath}   
\usepackage{amsfonts}    
\usepackage{amssymb}   
\usepackage{booktabs}
\usepackage{threeparttable}
\usepackage{subfigure}
\usepackage{comment}
\usepackage{xcolor}
\usepackage[normalem]{ulem}
\newcommand\redsout{\bgroup\markoverwith{\textcolor{red}{\rule[0.5ex]{4pt}{2pt}}}\ULon}
\newtheorem{theorem}{\textbf{Theorem}}

\newtheorem{lemma}{\textbf{Lemma}}

\newtheorem{remark}{\textbf{Remark}}
\newtheorem{example}{\textbf{Example}}
\newtheorem{assumption}{\textbf{Assumption}}

\DeclareMathOperator{\rank}{rank}
\DeclareMathOperator{\diag}{diag}
\DeclareMathOperator{\cond}{cond}
\DeclareMathOperator{\tr}{tr}
\newcommand{\HB}[1]{{\color{red}#1}}

\newcommand*{\QEDA}{\hfill\ensuremath{\blacksquare}}   
\allowdisplaybreaks[1]
 
\IEEEoverridecommandlockouts                              

\title{\LARGE \bf
	Model-Free Optimal Control of Linear Multi-Agent Systems via Decomposition and Hierarchical Approximation}

\author{Gangshan Jing,~He Bai,~Jemin George~and~Aranya Chakrabortty
	\thanks{G.~Jing and A. Chakrabortty are with  North Carolina State University, Raleigh, NC 27695, USA.
		{\tt\small gjing@ncsu.edu; achakra2@ncsu. edu}}%
	\thanks{H.~Bai is with Oklahoma State University, Stillwater, OK 74078, USA.
		{\tt\small he.bai@okstate.edu}}%
	\thanks{J.~George is with the U.S. Army Research Laboratory, Adelphi, MD 20783, USA.
		{\tt\small jemin.george.civ@mail.mil}}%
}

\allowdisplaybreaks[3]

\begin{document}

	\maketitle

	\begin{abstract}
		Designing the optimal linear quadratic regulator (LQR) for a large-scale multi-agent system (MAS) is time-consuming since it involves solving a large-size matrix Riccati equation. The situation is further exasperated when the design needs to be done in a model-free way using schemes such as reinforcement learning (RL). To reduce this computational complexity, we decompose the large-scale LQR design problem into multiple smaller-size LQR design problems. We consider the objective function to be specified over an undirected graph, and cast the decomposition as a graph clustering problem. The graph is decomposed into two parts, one consisting of independent clusters of connected components, and the other containing edges that connect different clusters. Accordingly, the resulting controller has a hierarchical structure, consisting of two components. The first component optimizes the performance of each independent cluster by solving the smaller-size LQR design problem in a model-free way using an RL algorithm. The second component accounts for the objective coupling different clusters, which is achieved by solving a least squares problem in one shot. Although suboptimal, the hierarchical controller adheres to a particular structure as specified by inter-agent couplings in the objective function and by the decomposition strategy. Mathematical formulations are established to find a decomposition that minimizes the number of required communication links or reduces the optimality gap. Numerical simulations are provided to highlight the pros and cons of the proposed designs.
	\end{abstract}
	
	\vspace{-5mm}
	\section{Introduction}
	
The classical linear quadratic regulator (LQR) problem \cite{Anderson07}  has been solved by a variety of approaches such as Kleinman's iterative algorithm \cite{Kleinman68}, semi-definite programming \cite{Balakrishnan03}, and policy gradient algorithms \cite{Bu20}. When applied to large-scale multi-agent systems (MASs), however, these methods often result in poor numerical performance due to the computation of extremely large-dimensional gain matrices. In other words, they produce encouraging results when run {\it offline}, but their execution becomes time-taking and numerically expensive when {\it real-time} control actions need to be taken. The problem becomes even more complex when the MAS model is unknown to the designer and data-driven techniques \cite{sutton}-\cite{ben} such as reinforcement learning (RL) need to be used. For instance, various RL algorithms such as actor-critic methods \cite{Kiumarsi17}, Q-learning \cite{meyn}, policy gradient algorithms \cite{Fazel18}, integral concurrent learning \cite{dixon}, and adaptive dynamic programming (ADP) \cite{Powell07}-\cite{Jing20} have been proposed for solving model-free LQR, but all of them involve large-size matrix inversions at every iteration of the learning phase, which drastically increases computational complexity, resulting in long learning times. The controller is usually centralized in  implementation, and even if it is distributed \cite{Borrelli08}-\cite{Li19} it comes at the cost of specific structural assumptions, or restrictive conditions on the cost function and the agent dynamics.

In this paper we propose a {\it hierarchical} RL scheme to resolve this computational bottleneck of conventional RL-based LQR. The hierarchy follows from decomposing a high-dimensional LQR problem into several smaller-sized LQR problems.  We consider a MAS where each agent has its own decoupled linear dynamics. The state penalty term in the control objective (i.e., the $Q$ matrix in the LQR cost function) couples these dynamics through a connected, undirected graph. The decomposition problem is posed as a clustering problem for this graph. The graph is decomposed into two parts, one consisting of multiple decoupled clusters of connected components, and the other containing edges that connect the different clusters. Accordingly, the proposed RL controller is composed of two components, one optimizing the performance of each decoupled cluster by solving a smaller-sized independent LQR design, and the other accounting for the performance coupling the different clusters by solving a least squares problem in one shot. 

The two main benefits of this hierarchical strategy are that (i) it drastically reduces learning time, and that (ii) the resulting controller inherits a special structure from the graph embedded in the $Q$ matrix as well from the decomposition strategy that enables it to be implemented in a cluster-wise distributed fashion, requiring far less number of communication links than conventional LQR.  We formulate a mixed-integer quadratic program (MIQP) to minimize the number of communication links. Because of the  decomposition the controller is suboptimal, however. We, therefore, present a minimum cut graph partitioning approach by which one can reduce the optimality gap. Compared to the conventional structure-constrained distributed LQR designs reported in the literature such as in \cite{Borrelli08,Nguyen16}, our cost function and agent models are much more generic.

Hierarchical optimal control has been studied in papers such as \cite{Nguyen16,Chow76} (model-based) and \cite{sayakcdc} (model-free) under the assumption of time-scale separation in the plant dynamics. A related method based on spectral separation of the controllability Gramian was also recently proposed in \cite{sadamoto}. These methods, when studied in the context of a MAS, often translate into clustering of agents \cite{Monshizadeh14,Cheng20}. The clustering in our problem, in contrast, is not on the agents but rather on the graph that defines the control objective for the agents. Our design is also fundamentally different from most cooperative control problems in the MAS literature such as \cite{Ren08,Bai11}, where the objective is to \textit{stabilize} the agents to a desired equilibrium set that describes a collective behavior. Although such stabilization objective may be characterized by an optimal control formulation \cite{Movric13}, the cost functions in those cases are not predefined, and have to satisfy specific conditions. We illustrate the effectiveness of our designs using a classical example from formation maneuver control. 

Some preliminary results on this design have been reported in our recent conference paper \cite{acc2020}. Compared to those results, the problem formulation as well as the control designs in this paper are significantly more extensive and in-depth. Stability proofs, solution of MIQP for optimal decomposition, and graph partitioning for minimization of the optimality gap are also completely new contributions. 
	
The rest of the paper is organized as follows. Section \ref{sec: problem formulation} formulates the model-free optimal control problem for linear heterogeneous MAS. Section \ref{sec:Het} proposes the hierarchical control framework via decomposition. Section \ref{sec: optimal decomposition} analyzes the influence of the decomposition on the number of communication links and the performance gap, and presents an MIQP problem to find an optimal decomposition strategy. Section \ref{sec: application} presents numerical results on formation maneuver control. Section \ref{sec: conclusion} concludes the paper. Stability proofs and other theoretical results are presented in the appendix.

	\textbf{Notation}: Throughout the paper, $\mathcal{G}=(\mathcal{V},\mathcal{E})$ denotes an undirected graph with $N$ vertices, where $\mathcal{V}=\{1,...,N\}$ is the set of vertices, $\mathcal{E}\subset\mathcal{V}\times\mathcal{V}$ is the set of edges; $\mathbf{0}_{p\times q}$ denotes the $p$ by $q$ zero matrix; $I_d$ is the $d\times d$ identity matrix; $e_i\in\mathbb{R}^N$ is a $N$-dimensional unit vector with the $i$-th element being 1 and other elements being 0's; $\otimes$ denotes the kronecker product. Define $\diag\{A_1,...,A_N\}$ as a block diagonal matrix with $A_i$'s on the diagonal. Given a matrix $A\in\mathbb{R}^{n\times n}$, $\lambda_i(A)$ denotes the $i$-th eigenvalue of $A$ such that $\lambda_1(A)\leq...\leq\lambda_n(A)$, $\lambda_{\min}(A)=\lambda_1(A)$, $\lambda_{\max}(A)=\lambda_n(A)$, $A\succ0$ implies that $A$ is positive definite, $A\succeq0$ implies that $A$ is positive semi-definite, and $\text{tr}(A)$ denotes the trace of $A$; Similarly, $\sigma_1(A)\leq...\leq\sigma_n(A)$ are singular values of $A$, $\sigma_{\min}(A)=\sigma_1(A)$, $\sigma_{\max}(A)=\sigma_n(A)$. We use $\cond(A)=\sigma_{\max}(A)/\sigma_{\min}(A)$ to denote the condition number of matrix $A$. For matrices $A,B\succeq0$, $A\leq B$ implies that $B-A\succeq0$, $A\geq B$ if $B\leq A$. Given a scalar $x$, $\lfloor x\rfloor$ denotes the largest integer smaller than or equal to $x$.
	
	\vspace{-3mm}
	\section{Problem Formulation}\label{sec: problem formulation}
	Consider a MAS composed of $N$ agents. Each agent $i$ is a linear time-invariant system described as
	\begin{equation}\label{MAS}
	\dot{x}_i=A_ix_i+B_iu_i, ~~~~i=1,...,N
	\end{equation}
	where $x_i\in\mathbb{R}^n$ and $u_i\in\mathbb{R}^m$ are the state and control input of agent $i$. Throughout the paper, we will consider $A_i$ and $B_i$ as unknown, but their dimensions are known. Let $x=(x_1^{\top} ,...,x_N^{\top} )^{\top} $, $u=(u_1^{\top} ,...,u_N^{\top} )^{\top} $, $\mathcal{A}=\diag\{A_1,...,A_N\}$, $\mathcal{B}=\diag\{B_1,...,B_N\}$. The overall model of the system is written in a compact form as
	\begin{equation}\label{compact system}
	\dot{x}=\mathcal{A}x+\mathcal{B}u.
	\end{equation}
	We assume that the agent dynamics are coupled through the cost function that needs to be minimized using a state-feedback LQR controller. Without loss of generality, we use an undirected graph $\mathcal{G}=(\mathcal{V},\mathcal{E})$ to describe the coupling relationship among the agents in the cost function. It is important to note that the graph $\mathcal{G}$ is not a communication or interaction graph between the agents. It is a graph that defines the coupling between the agent dynamics in the control objective. Similar settings have been considered in \cite{Borrelli08} and \cite{Nguyen16}. The cost function is given as
	\begin{equation}\label{original}
	J(x(0),u)=\int_0^\infty x^{\top} (\tau)Qx(\tau)+u^{\top} (\tau)Ru(\tau)\,\,d\tau,
	\end{equation}
	where $Q\succeq0$ and $R\succ0$ are in the following forms:
	\begin{equation}\label{Q and R}
	Q=\bar{Q}+G\otimes \tilde{Q},~~~~ R=\diag\{R_1,...,R_N\},
	\end{equation}
	with $\bar{Q}=\diag\{\bar{Q}_{11},...,\bar{Q}_{NN}\}$ representing the subsystem-level objective, $\bar{Q}_{ii}\in\mathbb{R}^{n\times n}$ and $\bar{Q}_{ii}\succeq0$. $R_i\in\mathbb{R}^{m\times m}$ and $R_i\succ0$ for $i=1,...,N$. $G=[G_{ij}]\in\mathbb{R}^{N\times N}$ is a Laplacian matrix corresponding to graph $\mathcal{G}=(\mathcal{V},\mathcal{E})$, i.e., $G_{ii}=\sum_{j=1}^N|G_{ij}|$ for $i=1,...,N$, and $G_{ij}<0$ if $(i,j)\in\mathcal{E}$ and $i\neq j$; $\tilde{Q}\in\mathbb{R}^{n\times n}$ and $\tilde{Q}\succeq0$. The second component in $Q$ represents the objective across the different subsystems. We observe that if $\mathcal{G}$ is not connected, then the optimal control problem can be decomposed immediately according to its independent connected components. Without loss of generality, we assume $\mathcal{G}$ is connected throughout the rest of the paper.
	
	Given the MAS (\ref{MAS}) with state $x$ measurable, we would ideally like to find the controller $u^*$ such that the performance index (\ref{original}) with matrices $Q$ and $R$ defined in (\ref{Q and R}) is minimized despite $\mathcal{A}$ and $\mathcal{B}$ being unknown. We make the following assumption to guarantee existence and uniqueness of the optimal controller.
	\begin{assumption}\label{as control and observe}
The pair $(\mathcal{A},\mathcal{B})$ is controllable and $(Q^{1/2},\mathcal{A})$ is observable. 
	\end{assumption}

	For heterogeneous MAS with $n\geq2$, $Q\succ0$ is sufficient but not necessary for observability of $(Q^{1/2},\mathcal{A})$. However, if $n=1$, then $(Q^{1/2},\mathcal{A})$ is observable only if $Q\succ0$. When $\mathcal{A}$ and $\mathcal{B}$ are known, the optimal controller for system (\ref{compact system}) to minimize (\ref{original}) is $u=-Kx=-R^{-1}\mathcal{B}^{\top}Px$ (see \cite{Anderson07}), where $P$ is the solution of the following algebraic Riccati equation:
	\begin{equation}\label{ARE}
	P\mathcal{A}+\mathcal{A}^\top P+Q-P\mathcal{B}R^{-1}\mathcal{B}^\top P =0.
	\end{equation}
	For MAS (\ref{MAS}), $K\in\mathbb{R}^{mN\times nN}$ can be partitioned into $N^2$ blocks $K(i,j)\in\mathbb{R}^{m\times n}$. Accordingly, the control input for agent $i$ can be written as $u_i=\sum_{j=1}^NK(i,j)x_j$. We define the communication graph associated with controller $u$ as   $\mathcal{G}_c(u)=(\mathcal{V},\mathcal{E}_c(u))$, where 
	\begin{equation}\label{E_c}
	\mathcal{E}_c(u)=\{(i,j)\in\mathcal{V}\times\mathcal{V}: K(i,j)\neq\mathbf{0}_{m\times n}\}.
	\end{equation}
	Therefore, communication between agents $i$ and $j$ is not required if $(i,j)\notin\mathcal{E}_c$. In the model-free case, where $\mathcal{A}$ and $\mathcal{B}$ are unknown, equation (\ref{ARE}) can be transformed into an equation that is independent of $\mathcal{A}$ and $\mathcal{B}$, and is based on the measurement of the states and the control inputs \cite{Vrabie09,Jiang12,Kiumarsi17,Jing20}. The matrix $P$ can then be obtained by implementing an RL algorithm, which solves a least squares problem at each iteration.
	
Both the model-based and model-free control methods, however, are  time-consuming if the MAS is of large scale. The main purpose of this paper is to provide an alternative approach for synthesizing the controller based on a decomposition of the cost function via clustering of the graph $\mathcal G$ that can reduce learning time significantly. This decomposition is described in the next section.
	
	\begin{remark}
	The state vector of every agent is assumed to have the same dimension $n$ for simplicity, but all of our designs and analysis in the forthcoming sections also apply to the case when agents have different dimensional states and control inputs under certain settings. More specifically, let $x_i\in\mathbb{R}^{n_i}$ and $u_i\in\mathbb{R}^{m_i}$. We introduce $C_i\in\mathbb{R}^{n\times n_i}$ with $n\leq n_i$ and $\rank(C_i)=n$ for $i=1,...,N$. The cost function can be reformulated as $J(x(0),u)=\int_0^\infty x^{\top}(\tau)C^{\top} QCx(\tau)+\sum_{i=1}^Nu_i^{\top} (\tau)R_iu_i(\tau)\,\,d\tau$, where $C=\diag\{C_1,...,C_N\}$, $Q$ is still defined in (\ref{Q and R}). 
	\end{remark}

		\vspace{-0.3cm}
	\section{Hierarchical Control for MAS}\label{sec:Het}
	
	To reduce learning time, we propose a hierarchical RL design that seeks a suboptimal controller for the general case when the agents have non-identical dynamics. Throughout the section, for ease of understanding the RL design, we will assume that the decomposition strategy is given, implying that the agents have already been decomposed into multiple clusters. The actual ways of developing this decomposition will be presented in Section IV. 
	
	\vspace{-3mm}
	\subsection{Hierarchical Control Framework}\label{subsec: dec for Het}
	
	Suppose that the vertex set $\mathcal{V}$ is decomposed into $s\leq N$ disjoint vertex sets $\mathcal{V}_j$, $j=1,...,s$. The $j$-th cluster is composed of $N_j$ nodes $\{j_1,..., j_{N_j}\}$. Let $n_j=nN_j$ and $m_j=mN_j$. The dynamics of the $j$-th cluster is written as
	\begin{equation}\label{clique system}
	\dot{\mathbf{x}}_j=\mathcal{A}_j\mathbf{x}_j+\mathcal{B}_j\mathbf{u}_j, ~~~~j=1,...,s,
	\end{equation}
	where $\mathbf{x}_j=(x_{j_1}^{\top} ,...,x_{j_{N_j}}^{\top} )^{\top} \in\mathbb{R}^{n_j}$, $\mathbf{u}_j=(u_{j_1}^{\top} ,...,u_{j_{N_j}}^{\top} )^{\top} \in\mathbb{R}^{m_j}$,  $\mathcal{A}_j=\diag\{A_{j_1},...,A_{j_{N_j}}\}\in\mathbb{R}^{n_j\times n_j}$, $\mathcal{B}_j=\diag\{B_{j_1},..., B_{j_{N_j}}\}\in\mathbb{R}^{n_j\times m_j}$. Without loss of generality, we assume that the agents belonging to the same cluster have contiguous indices, i.e., $j_{i+1}=j_i+1$ and $(j+1)_1=j_{N_j}+1$. In the case when the indices of agents are not in such an order, one can relabel them to make this condition satisfied. The matrix $G$ can then be decomposed as 
	\begin{equation}\label{decompose G}
	G=G_1+G_2,
	\end{equation}
	where $G_1$ is a block-diagonal Laplacian matrix with $s$ blocks and each block corresponds to a subgraph of  $\mathcal{G}$, and $G_2$ is a Laplacian matrix where $G_2(i,i)=\sum_{j=1}^N|G_2(i,j)|$ that describes the couplings between the different clusters. Then both $G_1$ and $G_2$ are positive semi-definite. 
	
	Given the decomposition $\mathcal{V}=\mathcal{V}_1\cup...\cup\mathcal{V}_s$, the graph $G_2$ is constructed as follows:
	
	Step 1. Let $G_2=G$. Replace each block on the diagonal of $G_2$ corresponding to each cluster $\mathcal{V}_j$ by a zero block. The remaining off-diagonal elements remain unchanged.
	
	Step 2. Replace the diagonal elements of $G_2$ by $G_2(i,i)=\sum_{j=1}^N|G_2(i,j)|$.
	
\begin{figure}
		\centering
		\includegraphics[width=8cm]{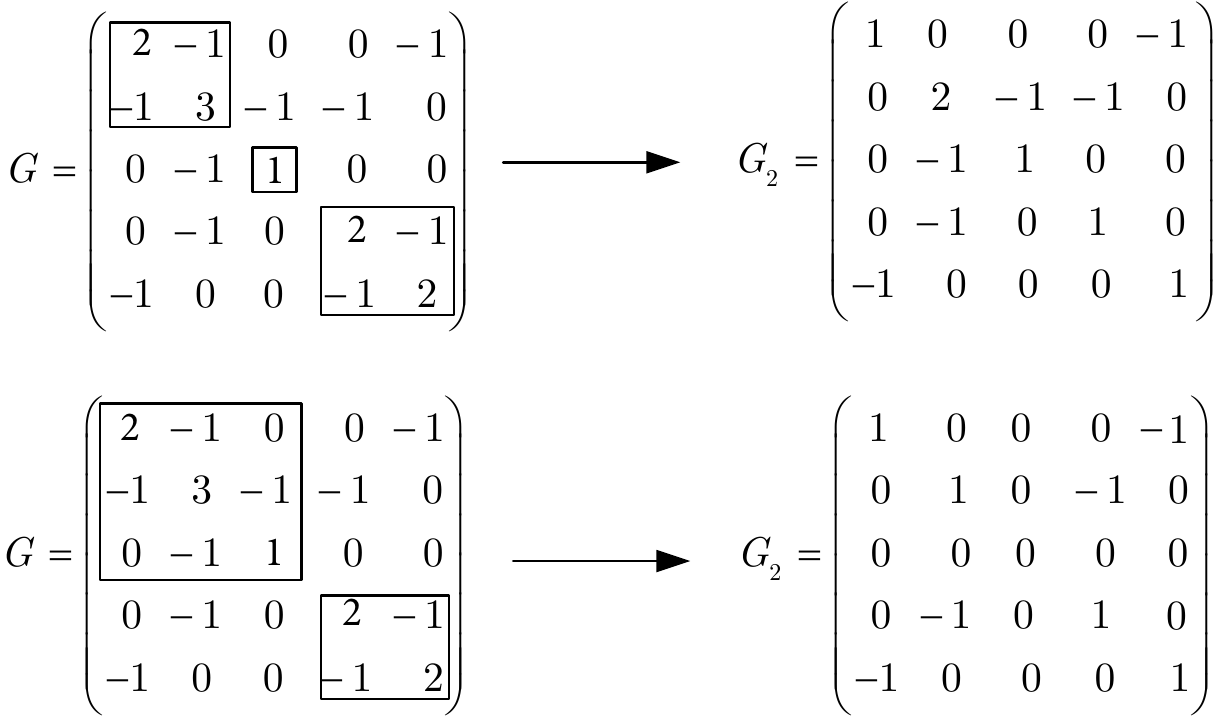}
		\caption{Two examples for the decomposition of $G$. In the first example, the agents are decomposed to three clusters $\{1,2\}$, $\{3\}$ and $\{4,5\}$. In the second example, the agents are decomposed to two clusters $\{1,2,3\}$ and $\{4,5\}$.} \label{fig decomposition example}
		\vspace{-0.7cm}	
	\end{figure}

	Once $G_2$ is constructed, $G_1$ will be determined accordingly. In Fig. \ref{fig decomposition example}, we give two examples to demonstrate two different decompositions for the same $G$. 
	Similar to $G$, the four matrices $Q$, $\bar{Q}$, $\tilde{Q}$ and $R$ can also be transformed corresponding to the new indices of the agents. For simplicity of notation, we assume that $Q$, $\bar{Q}$, $\tilde{Q}$ and $R$ correspond directly to the relabelled agents. From (\ref{Q and R}), we have

	\begin{equation}
	Q=\bar{Q}+(G_1+G_2)\otimes \tilde{Q}.
	\end{equation}
	We define $\hat{Q}=\bar{Q}+G_1\otimes \tilde{Q}$, 
	and write $\hat{R}_j=\diag\{R_{j_1},...,R_{j_{N_j}}\}\in\mathbb{R}^{m_j\times m_j}$ for the $j$-th cluster. Then (\ref{Q and R}) can be written as 
	\begin{equation}
	Q=\hat{Q}+G_2\otimes \tilde{Q},~~~~R=\diag\{\hat{R}_1,...,\hat{R}_s\}
	\end{equation}
	where $\hat{Q}=\diag\{\hat{Q}_1,..., \hat{Q}_s\}$. Accordingly, the global control objective in \eqref{original} can be decomposed into group-level objectives $J_j$, $j=1,\ldots,s$ and network-level objective $J_{\mathcal{G}}$ as 
	\begin{align}\label{Jdecom}
	    J(x(0),u)= \sum_{j=1}^s J_j(\mathbf{x}_j(0),\mathbf{u}_j) + J_{\mathcal{G}}(x(0),u),
 	\end{align}
 	where
	\begin{align}\label{Jlocal}
	    J_j(\mathbf{x}_j(0),\mathbf{u}_j) &= \int_{0}^{\infty}\,\mathbf{x}_j^\top(\tau)  \hat{Q}_j \mathbf{x}_j(\tau) + \mathbf{u}_j^\top(\tau)  \hat{R}_j \mathbf{u}_j(\tau) \, d\tau,
	\end{align}
	and
	\begin{align}\label{Jglobal}
	    J_{\mathcal{G}}(x(0),u) &= \int_0^\infty x^{\top} (\tau) \left( G_2\otimes \tilde{Q} \right) x(\tau)\,\,d\tau.
	\end{align}
	The algebraic matrix Riccati equation corresponding to the group-level integral quadratic cost in \eqref{Jlocal} is
	\begin{equation}\label{locric}
	\mathcal{P}_j\mathcal{A}_j + \mathcal{A}_j^{\top} \mathcal{P}_j + \hat{Q}_j - \mathcal{P}_j\mathcal{B}_j\hat{R} ^{-1}_j\mathcal{B}_j^{\top} \mathcal{P}_j=0, \;j=1,...,s.
	\end{equation}

	From Assumption \ref{as control and observe}, $(\mathcal{A}_j,\mathcal{B}_j)$ is controllable. However, $(\hat{Q}_j^{1/2},\mathcal{A}_j)$ may not be observable for each $j=1,...,s$. To guarantee existence and uniqueness of the solution to~\eqref{locric}, we make the following assumption:

	\begin{assumption}\label{as Qj observable}
		$(\hat{Q}_j^{1/2},\mathcal{A}_j)$ is observable for $j=1,..., s$.
	\end{assumption}
	
	\begin{remark}
		For a general matrix $\mathcal{A}_j$, Assumption \ref{as Qj observable} may be satisfied even when $\hat{Q}_j$ is singular. Observability of $(Q^{1/2},\mathcal{A})$ and $(\hat{Q}_j^{1/2},\mathcal{A}_j)$ can both be guaranteed without knowing the dynamics of each agent if $\hat{Q}_j\succ0$ for all $j=1,...,s$.
	\end{remark}
	
	The decomposition of the global control objective as given in \eqref{Jdecom} allows the individual groups to solve for their local optimal control gains (that minimizes \eqref{Jlocal}) in parallel. Next, we present an approximate control to account for the coupled objective given in \eqref{Jglobal}.  Motivated by \cite{Nguyen16}, we define
	\begin{equation}
	\mathcal{R}^{-1} = {R}^{-1} + \tilde{R}, \label{rtil}
	\end{equation}
	where the expression for $\tilde{R}$ will be derived shortly. Replacing ${R}^{-1}$ with $\mathcal{R}^{-1}$ in \eqref{ARE}
	yields
	\begin{align*}\label{block ARE}
	&\mathcal{P}\mathcal{A}+\mathcal{A}^\top\mathcal{P}+Q-\mathcal{P}\mathcal{B}\mathcal{R}^{-1}\mathcal{B}^\top\mathcal{P} \nonumber \\
	&= \underbrace{\mathcal{P}\mathcal{A}+\mathcal{A}^\top\mathcal{P} + \hat{Q} -
	\mathcal{P}\mathcal{B}{R}^{-1}\mathcal{B}^\top\mathcal{P}}_{\text{decoupled part}} + G_2 \otimes \tilde{Q} - \mathcal{P}\mathcal{B}\tilde{R}\mathcal{B}^\top\mathcal{P} \nonumber  \\
	&= \diag\{\mathcal{P}_j\mathcal{A}_j + \mathcal{A}_j^{\top}\mathcal{P}_j + \hat{Q}_j - \mathcal{P}_j\mathcal{B}_j\hat{R}_j^{-1}\mathcal{B}_j^{\top} \mathcal{P}_j\}\\ 
	& ~~~+ G_2 \otimes \tilde{Q} - \mathcal{P}\mathcal{B}\tilde{R}\mathcal{B}^\top\mathcal{P},
	\end{align*}
	indicating that the Riccati equation can be decomposed into multiple decoupled smaller-sized Riccati equations if 
	\begin{equation}\label{LQRP}
	G_2 \otimes \tilde{Q} - \mathcal{P}\mathcal{B}\tilde{R}\mathcal{B}^\top\mathcal{P}=0. 
	\end{equation}
	Therefore, we select $\tilde{R}$ as the solution of (\ref{LQRP}), where the block diagonal matrix $\mathcal{P} = \diag\{\mathcal{P}_1,\ldots,\mathcal{P}_s\}$ is obtained by solving the set of $s$ decoupled Riccati equations given in \eqref{locric}. The controller is designed hierarchically as 
	\begin{equation}\label{hierarchcial controller}
	u_h = -\mathcal{R}^{-1}\mathcal{B}^\top\mathcal{P}x= \underbrace{-R^{-1}\mathcal{B}^\top\mathcal{P}x}_{\text{local}}\underbrace{-\tilde{R}\mathcal{B}^\top\mathcal{P}x}_{\text{global}},
	\end{equation}
	where the first term is the local controller that can be obtained by solving multiple decoupled smaller-size Riccati equations, all in parallel, and the second term is the global controller based on $\tilde{R}$ solved from (\ref{LQRP}).

	However, it may not always be possible to find a $\tilde{R}$ satisfying \eqref{LQRP}. If $\mathcal B$ is square and non-singular (i.e., each agent is a fully actuated system), then $\tilde{R}$ follows simply as $\tilde{R}= (\mathcal{P}\mathcal{B})^{-1}(G_2\otimes\tilde{Q})(\mathcal{B}^{\top}\mathcal{P})^{-1}$. However, this expression does not hold when $\mathcal B\in\mathbb{R}^{nN\times mN}$ is rectangular and thus not invertible. In that case one may obtain $\tilde{R}$ by solving the following least squares semidefinite program (LSSDP):
	\begin{equation}\label{lssdp}
	\begin{split}
	&\min_{\tilde{R}} ||\mathcal{P}\mathcal{B}\tilde{R}\mathcal{B}^\top\mathcal{P} -G_2 \otimes \tilde{Q} ||_F\\
	\mbox{s.t}.&~~~~~~~~~\tilde{R}\succeq0.
	\end{split}
	\end{equation}		
	Let $(\mathcal{P}\mathcal{B})^+$ be the Moore-Penrose inverse of $\mathcal{P}\mathcal{B}$. In \cite{Penrose56}, it was shown that the solution $(\mathcal{P}\mathcal{B})^+(G_2\otimes\tilde{Q}){(\mathcal{P}\mathcal{B})^\top}^{+}$ has minimum norm among all solutions to (\ref{lssdp}). Therefore, one may compute $\tilde{R}$ as
	\begin{equation}\label{Rtilde}
	\tilde{R}=(\mathcal{P}\mathcal{B})^+(G_2\otimes\tilde{Q}){(\mathcal{P}\mathcal{B})^\top}^{+}.
	\end{equation}
	Specifically, when $\mathcal B\in\mathbb{R}^{nN\times mN}$ is full column rank, the least square solution is uniquely determined by
	$$\tilde{R}=  \left( (\mathcal{P}\mathcal{B})^\top\mathcal{P}\mathcal{B} \right)^{-1} (\mathcal{P}\mathcal{B})^\top \left(G_2 \otimes \tilde{Q}\right) \mathcal{P}\mathcal{B} \left((\mathcal{P}\mathcal{B})^\top\mathcal{P}\mathcal{B} \right)^{-1}.
	\label{ls}$$
	Rewrite $\mathcal{P}=\diag\{p_1^{\top},...,p_N^{\top}\}^{\top}$, where $p_i\in\mathbb{R}^{n\times nN}$. From (\ref{hierarchcial controller}), the controller for each agent $i$ can be written as:
	\begin{equation}\label{controller for i}
	u_h^{(i)}=-R_i^{-1}B_i^{\top}\mathcal{P}_{j^i,i}\mathbf{x}_{j^i}-\sum_{k=1}^N\tilde{R}_{ik}B_k^{\top}p_kx,
	\end{equation}
	where $j^i$ specifies the cluster containing $i$. $\mathcal{P}_{j^i,i}\in\mathbb{R}^{n\times nN_{j^i}}$ is a submatrix consisting of the continuous $n$ rows of $\mathcal{P}_{j^i}$ corresponding to agent $i$. Since $\tilde{R}$ inherits the structure of $G_2$ and $\mathcal{P}$ is block-diagonal, there are many zero entries in $\tilde{R}$ and $p_k$. As a result, agent $i$ only requires information of agents from its own cluster and other clusters for which the corresponding entries in $\tilde R$ are non-zero. Theorem 2 in Subsection \ref{subsec: decom and comm} characterizes what entries of $\tilde R$ are zero given a $G_2$. Moreover, the controller (\ref{controller for i}) can be implemented in a decentralized way, as discussed  in Remark \ref{re distributed learning}.
	
	Instead of minimizing the original objective function \eqref{original}, the approximate controller \eqref{hierarchcial controller} minimizes
	\begin{equation}\label{app}
	\mathcal{J}(x(0),u) = \int_0^\infty x^\top \mathcal{Q}x + u^\top \mathcal{R} u \,\, dt, 
	\end{equation}
	where $\mathcal{Q}=\hat{Q}+\mathcal{P}\mathcal{B}\tilde{R}\mathcal{B}^\top \mathcal{P}$, and $\mathcal R$ follows from \eqref{rtil}. The following theorem shows that the hierarchical controller (\ref{hierarchcial controller}) is stabilizing. The proof is given in Section \ref{appendix}.
	\begin{theorem}\label{th stability}
		Under Assumptions \ref{as control and observe} and \ref{as Qj observable}, the hierarchical controller (\ref{hierarchcial controller}) guarantees that the MAS (\ref{MAS}) is globally asymptotically stable. \QEDA
	\end{theorem}

	\subsection{Hierarchical RL Algorithm}
	
	We next use \cite[Algorithm 3]{Jing20}\footnote{Although \cite{Jing20} focuses on minimum-cost variance control, the proposed algorithms can be applied to conventional LQR problems in a specific case, see \cite[Remark 1]{Jing20}.} to find the decomposed LQR controllers in a model-free way. The matrix $\mathcal{B}$ is needed for solving the least squares problem (\ref{LQRP}). For our design $\mathcal{B}$ can be estimated by estimating  $\mathcal{B}_j$ for each cluster $j$ at the very first step of the RL algorithm as shown in \cite{Jing20}. Suppose that the cost function is decomposed into $s$ clusters, and that Assumption \ref{as Qj observable} is satisfied. Matrices $G_1$ and $G_2$ are determined accordingly. By combining the RL algorithm in \cite{Jing20} with our hierarchical control framework, we propose Algorithm \ref{alg:2} as the final RL algorithm for heterogeneous MAS.
	
	\begin{algorithm}[h]
		\caption{RL Algorithm for Optimal Control of Heterogeneous MAS via Hierarchical Control}\label{alg:2}
		\textbf{Input}: $\hat{Q}_j$, $j=1,...,s$, $G_2\otimes \tilde{Q}$, $R=\diag\{\hat{R}_1,...,\hat{R}_s\}$.\\
		\textbf{Output}: Optimal controller $u^*$
		\begin{itemize}
			\item[1.] Run \cite[Algorithm 3]{Jing20} to solve the LQR problem \eqref{Jlocal} for each cluster $j$, $j=1,...,s$, with group dynamics (\ref{clique system}). Obtain $\mathcal{P}_j$ and estimated $\mathcal{B}_j$ for the $j$-th LQR problem.
			\item[2.] Compute the Moore-Penrose inverse of $\mathcal{P}_j\mathcal{B}_j$ for $j=1,...s$. Then compute $\tilde{R}$ by (\ref{Rtilde}).     
			\item[3.] The hierarchical optimal controller is 
			$$u_h=-(R^{-1}\mathcal{B}^{\top}\mathcal{P}+\tilde{R}\mathcal{B}^{\top}\mathcal{P})x.$$	
		\end{itemize}
	\end{algorithm}	
	
	\begin{remark}\label{re distributed learning}
		Algorithm \ref{alg:2} is presented from the viewpoint of centralized learning. It can also be implemented in a decentralized way under specific conditions. One such condition is as follows. Consider that a {\it coordinator} is chosen from the agents in each cluster. The coordinator collects and transmits state information from and to all agents in that cluster. Communication between different clusters happens through these coordinators. The interaction relationship between different coordinators is determined by the structure of $\tilde{R}$ (details are shown in Theorem 2). Through communication, the coordinator in each cluster $j$ has access to $\mathbf{x}_j$ and $\mathbf{x}_i$ from any adjacent cluster $i$, as well as to  $\mathcal{P}_j\mathcal{B}_j$ and $\mathcal{P}_i\mathcal{B}_i$. Based on these information, each coordinator can use the steps of Algorithm 1 to learn the hierarchical optimal controller for cluster $j$ in a  decentralized fashion. Once the hierarchical control gain is learned, the controller (\ref{controller for i}) can be implemented in a decentralized way as well. In this scenario, coordinators in different clusters exchange state information of agents in their clusters. Then each agent is able to obtain from its coordinator the required state information of agents in other clusters. 
	\end{remark}
	
		\begin{remark}
		Not just model-free RL, the proposed hierarchical design is equally applicable to model-based optimal control. The strategy in that case will be to decompose the Riccati equation into multiple smaller-sized Riccati equations.
	\end{remark}

	\subsection{Numerical Comparison with Conventional RL}
	We close this section by citing a numerical example that compares the conventional RL algorithm of \cite{Jiang12} with our hierarchical RL algorithm.
	
\begin{figure}
		\centering
		\includegraphics[width=8cm]{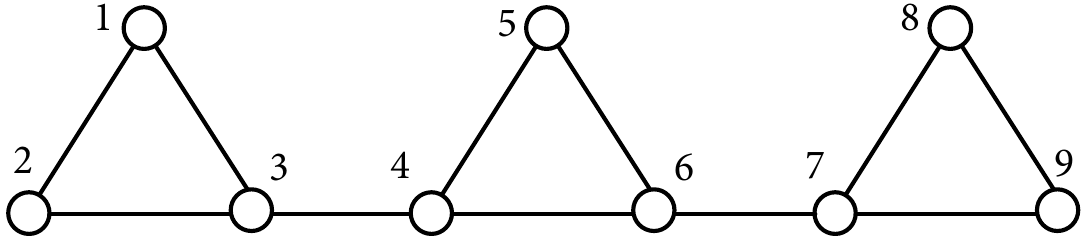}
		\caption{Graph $\mathcal{G}$ with 3 cliques, each clique contains 3 agents.} \label{fig clique graph}
	\end{figure}

	\begin{example} \label{ex compare with undecom}   
		Consider $\mathcal{G}$ as a graph composed of $s$ cliques. Each clique has $c$ agents, and will be referred to as a cluster. The graph between cliques is considered as an unweighted path graph. Moreover, there is only one link between any two connected cliques $i$ and $j$. See Fig. \ref{fig clique graph} as a demonstration of $\mathcal{G}$ for $s=c=3$. We set $G=L$, $\bar{Q}=0.5I_{nN}$, where $L$ is the Laplacian matrix, $Q=G\otimes I_n+\bar{Q}$ and $R=I_{mN}$. Then Assumption \ref{as Qj observable} always holds for arbitrary decomposition. For $n=4$ and $m=2$, we set $A_i=\begin{pmatrix}
		-I_2& I_2\\ \mathbf{0} &-\frac{i}{i+1}I_2
		\end{pmatrix}$ and $B_i=\begin{pmatrix}
		\mathbf{0}\\\frac{i}{i+1}I_2\end{pmatrix}$, $i=1,...,N$. For $n=8$ and $m=4$, we set $A_i=A_i\otimes I_2$ and $B_i=B_i\otimes I_2$. The component of each agent's initial state is randomly selected from $\{1,-1,0\}$.

		The computational time and the resulting performance index $J$ for each case are shown in Table \ref{tab2}. We see that the hierarchical RL algorithm saves a significant amount of time compared with the conventional RL algorithm. In Table \ref{tab2}, ``RL" denotes the conventional RL algorithm from \cite{Jiang12}, ``HRL" denotes Algorithm \ref{alg:2}, ``**" implies that the time for data collection is too long for conventional RL, ``OPT" is the optimal value by implementing the controller learned from conventional RL, ``SOP" means ``suboptimality", which is defined as $\frac{J_h-J^*}{J^*}$, where $J_h$ and $J^*$ are the performances of the hierarchical controller and the optimal controller, respectively. We find that as the problem size $Nn$ increases, closed-loop performance of the hierarchical controller becomes closer to optimal. This is because when there are more links in the overall graph, the links contained in $G_2$ become less important in the computation of the total performance index. For the conventional RL algorithm, the SOP for each case is close to zero.
	\end{example}  
	
	\begin{table}[htbp]	
		\centering
		\fontsize{9}{9}\selectfont
		\begin{threeparttable}
			\caption{Comparisons between different RL algorithms.}
			\label{tab2}
			\begin{tabular}{ccccccccc}
				\toprule			
				\multicolumn{4}{c}{Dimension}&\multicolumn{2}{c}{Time(sec)}&\multicolumn{3}{c}{Performance}\cr
				\cmidrule(lr){1-4} \cmidrule(lr){5-6}
				\cmidrule(lr){7-9}
				s&c&n&m&RL&HRL&OPT&HRL& SOP\cr
				\midrule
				3&2& 4& 2& 0.57&{\bf 0.07}& 28.76&{\bf 40.57}&41.08\%\cr
				3&3& 4& 2& 7.04&{\bf 0.08}& 52.75&{\bf 62.24}&18.00\%\cr
				3&4& 4& 2& 29.69&{\bf 0.24}& 81.03&{\bf 87.57}&8.08\%\cr
				4&4& 8& 4& **&{\bf 9.83}& 198.89&{\bf 209.29}&5.23\%\cr
				
				\bottomrule
			\end{tabular}
		\end{threeparttable}
	\end{table}
	
	In Example \ref{ex compare with undecom}, the agents corresponding to each clique are considered to form a cluster. There exist other ways to decompose the graph $\mathcal{G}$ in Fig. \ref{fig clique graph}. Given the number of clusters as three, the decomposition strategy used for this example is optimal in terms of the performance of the hierarchical controller. The reason will be shown in the next section. We will revisit this example in Subsection \ref{subsec: num for decomposition}.

	\section{Optimal Decomposition for Minimizing Communication and Suboptimality}\label{sec: optimal decomposition}
	
	To employ the hierarchical control framework presented in Section III, the decomposition strategy should be pre-specified. Given a MAS, there can be multiple ways to decompose $\mathcal G$ that may result in different performances and different amounts of control energy required. In this section, we analyze how a decomposition affects the communication graph $\mathcal{G}_c$ and the performance of the resulting hierarchical controller. Based on this analysis, we propose an approach to choose a decomposition minimizing the number of inter-agent communication links, or optimizing the performance of the hierarchical controller.
	
	\vspace{-4mm}
	\subsection{Relationship Between Decomposition and Communication Graph}\label{subsec: decom and comm}		
	
	Let $\tilde{R}\in\mathbb{R}^{mN\times mN}$ be the solution to (\ref{lssdp}). As $\mathcal B^\top \mathcal P$ is block-diagonal according to the clusters, the graph structure among different clusters is preserved in $\tilde R$ and $\tilde R \mathcal{B}^\top \mathcal P$. We partition $\tilde{R}$ into $s^2$ blocks $\tilde{R}(i,j)\in\mathbb{R}^{m_i\times m_j}$, and partition $G_2\in\mathbb{R}^{N\times N}$ into $s^2$ blocks $G_2(i,j)\in\mathbb{R}^{N_i\times N_j}$, $i,j=1,...,s$, respectively, according to the $s$ clusters. We state the following theorem.
	
	\begin{theorem}\label{th structure}
		Consider the MAS (\ref{MAS}) decomposed into $s$ clusters. For any two distinct clusters $i$ and $j$, if $G_2(i,j)=\mathbf{0}_{N_i\times N_j}$, then $\tilde{R}(i,j)=\mathbf{0}_{m_i\times m_j}$, and $\mathcal{E}_c(u_h)$ in (\ref{E_c}) contains no edges between the agents in cluster $i$ and in cluster $j$, where $u_h$ is the hierarchical controller  (\ref{hierarchcial controller}). \QEDA
	\end{theorem}

	This theorem indicates that our hierarchical controller (\ref{hierarchcial controller}) inherits the structure of $G_2$ for different clusters in the sense that any pair of agents from two different clusters do not need to interact with each other through the feedback  if there is no link between those two clusters in graph $\mathcal{G}$. Let $j\sim k$ and $j\nsim k$ denote the adjacent and non-adjacent relationships between any two clusters $j$ and $k$, respectively. The controller for agent $i$ in (\ref{controller for i}) can then be rewritten as
	\begin{equation}
	    u_h^{(i)}=-R_i^{-1}B_i^{\top}\mathcal{P}_{j^i,i}\mathbf{x}_{j^i}-\sum_{j^i\sim k}\tilde{R}_i(j^i,k)\mathcal{B}_k^{\top}\mathcal{P}_k\mathbf{x}_k,
	\end{equation}
	where $\tilde{R}_i(j^i,k)\in\mathbb{R}^{m\times mN_{j^i}}$ is a submatrix consisting of $m$ rows of $\tilde{R}(j^i,k)$ corresponding to agent $i$. 

	Consequently, the number of  communication links required for our hierarchical controller is upper bounded by $N^2-\kappa$, where  
	\begin{equation}
	\kappa=\sum_{i\nsim j}N_iN_j
	\end{equation}
	is the number of pairs of agents that do not need to communicate with each other.

	\vspace{-4mm}
	\subsection{Relationship Between Decomposition and Performance}\label{subsec: decom and perf}
	
	Let $u^*$ and $u_h^*$ be the optimal controllers corresponding to the original cost function $J(x(0),u)$ in (\ref{original}) and the approximate cost function $\mathcal{J}(x(0),u)$ in (\ref{app}), respectively. Here $\mathcal{J}(x(0),u)$ is a function of matrix $\mathcal{Q}$, which depends on the decomposition strategy. It can be verified that
	$$x^{\top}(0)Px(0)=J(x(0),u^*),$$
	$$x^{\top}(0)\mathcal{P}x(0)=\mathcal{J}(x(0),u_h^*),$$
	where $P$ and $\mathcal{P}$ are the solutions to the following two Riccati equations, respectively:
	\begin{equation}\label{original P}
	P\mathcal{A}+\mathcal{A}^{\top}P+Q-P\mathcal{B}R^{-1}\mathcal{B}^{\top}P=0,
	\end{equation}
	\begin{equation}\label{mathcalP}
	\mathcal{P}\mathcal{A}+\mathcal{A}^{\top}\mathcal{P}+\mathcal{Q}-\mathcal{P}\mathcal{B}(R^{-1}+\tilde{R})\mathcal{B}^{\top}\mathcal{P}=0.
	\end{equation}
	
	The following theorem shows the relationship between the optimal performance for the approximate problem, the optimal performance of the original problem, and the performance of the original problem with our approximate hierarchical controller.

	\begin{theorem}\label{th JJJ}
		Given MAS (\ref{MAS}) with initial state $x(0)$, the optimal value of the approximate cost (\ref{app}), the optimal value of the original cost (\ref{original}), and the value of the original cost for our hierarchical controller have the following relationship:
		\begin{equation}\label{three J}
		\mathcal{J}(x(0),u_h^*)\leq J(x(0),u^*)\leq J(x(0),u_h^*).
		\end{equation}
		\QEDA
	\end{theorem}		
	
	It is desirable to design a decomposition strategy such that the corresponding hierarchical controller $u_h^*$ makes $J(x(0),u_h^*)$ close to $J(x(0),u^*)$. To this end, we try to find the optimal decomposition such that $$\Delta J_h=J(x(0),u_h^*)-J(x(0),u^*)\geq0$$ is minimized.	Since the performance error always depends on the initial state $x(0)$, we propose to analyze the average performance and the expectation $\mathbb{E}(\Delta J_h)$  given a random initial state vector $x(0)$. The following theorem shows an explicit form of $\mathbb{E}(\Delta J_h)$.
	\begin{theorem}\label{th performance gap}
		Given MAS (\ref{MAS}), let $K_h=\mathcal{R}^{-1}\mathcal{B}^{\top}\mathcal{P}$ be the hierarchical control gain matrix, and $K=R^{-1}\mathcal{B}^{\top}P$ be the optimal control gain matrix. Suppose that the initial state vector $x(0)$ is a random variable with zero mean, and $\sigma^2I_{nN}$ ($\sigma>0$) as its covariance matrix\footnote{The assumption on the initial state here is to eliminate the dependence of the control performance on the initial conditions in our analysis for the relationship between decomposition and performance.  The original MAS optimal control problem is still deterministic.}. Then,
		\begin{equation}
		\mathbb{E}(\Delta J_h)=\sigma^2\tr(V),
		\end{equation}
		where $V$ is the solution of
		\begin{equation}\label{V equation}
		\mathcal{A}_s^{\top}V+V\mathcal{A}_s+W=0,
		\end{equation}
		where $\mathcal{A}_s=\mathcal{A}-\mathcal{B}\mathcal{R}^{-1}\mathcal{B}^{\top}\mathcal{P}$ and $W=(K_h-K)^{\top}R(K_h-K)$. \QEDA
	\end{theorem}

	Next, we present an upper bound for $\tr(V)$ depending on $\mathcal{P}$ and $G_2$, as stated in the following lemma. 
	
	\begin{lemma}\label{le bound for trace}	
		For the MAS given in (\ref{MAS}), the following statements hold for the matrices $V$ and $W$ defined in Theorem~\ref{th performance gap}:
		
		(i) If $\bar{Q}\succ0$, then 
		\begin{equation}\label{bound of traceU}
		\tr(V)\leq \frac{\lambda_{\max}(\mathcal{P}) \cond(\mathcal{P})\tr(W)}{\lambda_{\min}(\bar{Q})};
		\end{equation}
		
			(ii) \begin{equation}\label{bound of traceW}
			\tr(W)\leq f_1(\tr(G_2),\lambda_{\min}(\mathcal{P}))+f_2(\tr(G_2),\lambda_{\min}(\mathcal{P})),
			\end{equation}
			with			
			\begin{align}	
			f_1&=\frac{\tr^2(G_2)\tr^2(\tilde{Q})}{\lambda_{\min}^2(\mathcal{P})\sigma_l^2(\mathcal{B})}\lambda_{\max}(R)\label{f1},\\
			f_2&=\left[\lambda_{\max}(P)-\lambda_{\min}(\mathcal{P}) \right]\times\nonumber\\
			&\left[\tr(\mathcal{B}R^{-1}\mathcal{B}^{\top})+\frac{\sigma_{\max}^2(\mathcal{B})\tr(G_2)\tr(\tilde{Q})}{\lambda_{\min}^2(\mathcal{P})\sigma_l^2(\mathcal{B})}\right]\label{f2},
			\end{align}	
			where $\sigma_l(\mathcal{B})$ is the minimum nonzero singular value of $\mathcal{B}$.\QEDA	
	\end{lemma}	

	Lemma \ref{le bound for trace} implies that if there exists a decomposition such that compared with all other possible decomposition, $\tr(G_2)$ and $\lambda_{\max}(\mathcal{P})$ are minimal, and $\lambda_{\min}(\mathcal{P})$ is maximal, implying that $\cond(\mathcal{P})$ is minimal as well, then the upper bound of $\mathbb{E}(\Delta J_h)$ corresponding to this decomposition is also minimal. However, there may not exist a decomposition such that all these indices are optimized simultaneously. Since minimizing $\cond(\mathcal{P})$ is necessary for minimizing $\lambda_{\max}(\mathcal{P})$ and maximizing $\lambda_{\min}(\mathcal{P})$, we only consider $\tr(G_2)$ and $\cond(\mathcal{P})$ as the two most important indices for evaluating $\mathbb{E}(\Delta J_h)$.
	
	In conclusion, when $\bar{Q}\succ0$, we can view $\tr(G_2)$ and $\cond(\mathcal{P})$ as two of the most important factors affecting the performance of our hierarchical controller corresponding to any given decomposition.

		\begin{remark}
			From (\ref{locric}), $\mathcal{P}$ depends not only on $\hat{Q}$, but also on $\mathcal{A}$, $\mathcal{B}$ and $R$. For a heterogeneous MAS, if different agents have largely different dynamics and $R_i$ differs largely for different $i$, then these differences will strongly influence the resulting $\mathcal{P}$ for different decomposition strategies. Since the dynamics of the agents are unknown, we cannot utilize this information for  analyzing the relationship between the decomposition and the performance of the hierarchical controller. The performance is mainly determined by $\tr(G_2)$ and $\cond(\mathcal{P})$ only when the differences between the agent dynamics are small, and $R_i$ for different $i$ are similar.
		\end{remark}

	\subsection{Approaches for Optimizing Decomposition}

	Subsection \ref{subsec: decom and comm} shows that maximizing $\kappa=\sum_{i\nsim j}N_iN_j$ helps in reducing the number of required communication links in the hierarchical controller $u_h$. On the other hand, Subsection \ref{subsec: decom and perf} shows that minimizing $\tr(G_2)$ and $\cond(\mathcal{P})$ help optimize the performance of the closed-loop system. The index $\cond(\mathcal{P})$ is always influenced by $\mathcal{A}$, $\mathcal{B}$ and $R$, so it is impractical to optimize $\cond(\mathcal{P})$ by choosing the decomposition strategy without explicit knowledge of $\mathcal{A}$ and $\mathcal{B}$. However, minimizing $\tr(G_2)$ is independent of the system dynamics, and, therefore, is a tractable problem. Accordingly, we seek to maximize $\kappa$ for minimizing the number of communication links, and minimize $\tr(G_2)$ for optimizing the performance of our hierarchical controller.
	
	Given the desired number of clusters, minimizing $\tr(G_2)$ is equivalent to minimizing the total number of edges crossing any two different clusters, which is actually a minimum $s$-cut problem. In \cite{Goldschmidt94}, an algorithm is proposed to solve the minimum $s$-cut problem in polynomial time for a specified $s$. Maximizing $\kappa$, in comparison, is more complicated because the structure of the graph is important in computing $\kappa$. For example, in the first example shown in Fig. \ref{fig decomposition example}, $\kappa=2$ and $\tr(G_2)=6$, and the interaction between cluster 2 and cluster 3 is not required in the hierarchical controller. If we use another decomposition, say,  $\mathcal{V}_1=\{1\}$, $\mathcal{V}_2=\{2,3\}$ and $\mathcal{V}_3=\{4,5\}$, then $\kappa=0$ and $\tr(G_2)=6$, in which case, although the number of edges crossing the clusters remains the same, the resulting hierarchical control may require communications between two different clusters.
	
	Given $s$ as the desired number of clusters, we next find the decomposition for maximizing $\kappa$. Let $\eta_i\in\mathbb{R}^{N}$ be a vector where each component is either 0 or 1, $i=1,...,s$. We use $\eta_i$ to determine the members of cluster $i$. Let $\eta_i(k)$ be the $k$th element of $\eta_i$ and define
	
	$$\eta_i(k)=\left\{
	\begin{aligned}
	&1,~~\text{if}~k\in\mathcal{V}_i;\\
	&0,~~\text{otherwise},
	\end{aligned}
	\right.$$	
	where $\mathcal{V}_i$ is the set of agents in the $i$-th cluster. Then the number of agents in clusters $i$ is $N_i=|\mathcal{V}_i|=\eta_i^{\top}\mathbf{1}_N$. Given $s$ as the number of clusters, we use matrix $E=(\eta_1,...,\eta_s)\in\mathbb{R}^{N\times s}$ to denote a decomposition. Accordingly, $G_1(E)$ and $G_2(E)$ become the two decomposed matrices in (\ref{decompose G}) following this decomposition $E$, i.e., 
	\begin{equation}\label{decompose G E}
	G=G_1(E)+G_2(E).
	\end{equation}
	Note that $G_1(E)$ may not be block-diagonal but can be transformed to be so by relabelling the agents.

Next, we formulate a MIQP problem for seeking the decomposition maximizing $\kappa$. From the definition of $E$, we have $\kappa=\sum_{i\nsim j}N_iN_j=\sum_{i\nsim j}\eta_i^{\top}\mathbf{1}_N\mathbf{1}_N^{\top}\eta_j$. Let $\bar{G}=|G|$ where $\bar{G}_{ij}=|G_{ij}|$. For any two clusters $i$ and $j$, the total amount of interaction weights between them is $\langle \eta_i\eta_j^{\top},\bar{G}\rangle=\eta_i^{\top}\bar{G}\eta_j$. Therefore, $i\nsim j$ if and only if $\eta_i^{\top}\bar{G}\eta_j=0$. Let $l_{ij}\leq1-\eta_i^{\top}\bar{G}\eta_j/T$ and $\tau_{ij}=\eta_i^{\top}\mathbf{1}_N\mathbf{1}_N^{\top}\eta_j$, where $T$ is an integer greater than or equal to the sum of the weights of $\bar{G}$. Then $\eta_i^{\top}\bar{G}\eta_j/T=0$ if $i\nsim j$, and $\eta_i^{\top}\bar{G}\eta_j/T\leq1$ otherwise. If we consider $l_{ij}$ as a binary variable, then the maximum value of $l_{ij}$ is 1 if $i\nsim j$, and is 0 otherwise. Therefore, the decomposition that maximizes $\kappa$ is given by the solution to the following MIQP problem:
		\begin{equation}\label{MIQP}
		\begin{split}
		&~~~~~~~~~~~\min_{\eta_1,...,\eta_s} -\sum_{i=1}^s\sum_{j=1}^sl_{ij}\tau_{ij}\\
		\text{s.t.}~~ &~~~~~~~~~ \sum_{i=1}^s\eta_i=\mathbf{1}_N,~~ \eta_i^{\top}\mathbf{1}_N\geq1,\\
		&~~~~\tau_{ij}=\eta_i^{\top}\mathbf{1}_N\mathbf{1}_N^{\top}\eta_j, ~~l_{ij}\leq 1-\eta_i^{\top}\bar{G}\eta_j/T,\\
		&~~~~l_{ij}, ~\eta_i(k)\in\{0,1\},~~i,~j=1,..., s.
		\end{split}
		\end{equation}
The MIQP problem (\ref{MIQP}) can be  solved, for example, by branch and bound algorithms. Commercial software such as  Matlab, Lindo and  Gurobi can be used for this purpose, as will be shown in our simulations. We next present two properties.

\begin{theorem}\label{th simple case}
Let $\kappa^\ast(s)$ be the maximum value of $\kappa$ for $s$  clusters. Given matrix $G\in\mathbb{R}^{N\times N}$ in (\ref{Q and R}), $\kappa^*(s)$ is non-decreasing in $s$ and $\kappa^*(s)\leq z_0$, where $z_0$ is the number of zero elements in $G$.\QEDA
\end{theorem}	


Theorem \ref{th simple case} can be used to adjust $s$ to reduce the number of required communication links in the hierarchical controller. For example, in the second example of Fig. \ref{fig decomposition example}, $\kappa(E)=0    $ for any decomposition if $s=2$. Once we set $s=3$, there are two communication links that can be removed, as shown in the first example of Fig. \ref{fig decomposition example}. 

\vspace{-2mm}
\subsection{Numerical Verifications} \label{subsec: num for decomposition}

As we discussed in Subsection \ref{subsec: decom and perf}, when $\bar{Q}\succ0$, and $\cond(\mathcal{P})$ is similar for different decompositions, $\tr(G_2)$ is the key factor determining the performance of the hierarchical controller. In what follows, we will present two numerical examples where $\cond(\mathcal{P})$ is similar for different decompositions.

\begin{example}\label{ex compare decompositions}
	We reconsider the example with graph $\mathcal{G}$ in Fig. \ref{fig clique graph}. Matrices $Q$, $R$ and the agent dynamics are the same as those considered in Example \ref{ex compare with undecom}. By generating the initial states of agents 1000 times such that each time every component of the agent states is a random number chosen from a normal distribution with zero mean and variance 0.5, the average performance values for different decompositions are shown in Table \ref{tabhet2}. By solving a mixed integer linear program (MILP) formulation for the minimum $s$-cut problem, we obtain the second decomposition, which minimizes $\tr(G_2)$ and generates the hierarchical controller that has the best performance. By solving MIQP (\ref{MIQP}) using Gurobi in Matlab (total time taken is 7.8843 seconds), we obtain the third decomposition that maximizes $\kappa$, resulting in a controller that requires the fewest communication links.
	
	\begin{table}[htbp]	
		\centering
		\fontsize{7}{10}\selectfont
		\begin{threeparttable}
			\caption{Comparisons between different decompositions.}
			\label{tabhet2}
			\begin{tabular}{ccccccccc}
				\toprule			
				Decomposition   &    $\kappa$&$\tr(G_2)$&$\cond(\mathcal{P})$&$J$&$n_c$&SOP\\
				\midrule
				\{1,2\},\{3,..,7\},\{8,9\}& 4 & 8&16.4& 15.9&32&23.57\%\\
				\{1,2,3\},\{4,5,6\},\{7,8,9\}& 9& 4&17.1&14.3&27&10.20\%\\
				\{1,..,3\},\{4\},\{5,...,9\}& 15 & 6&17.0& 15.8&21&22.15\%\\
				Undecomposed& n/a & n/a&n/a&12.9&36&0\\
				\bottomrule
			\end{tabular}
		\end{threeparttable}
	\end{table}

\end{example}

	\begin{remark}
		When $\bar{Q}$ is well-conditioned (i.e., when $\cond(\bar{Q})$ is small), and $\lambda_{\max}(G)$ is not large, $\hat{Q}$ is well-conditioned because $\cond(\hat{Q})\leq(\lambda_{\max}(G_1)+\lambda_{\max}(\bar{Q}))/\lambda_{\min}(\bar{Q})$. Moreover, when all agents have identical dynamics and equal values of $R_i$, $\mathcal{P}$ will be well-conditioned as well. In that case, $\tr(G_2)$ is the most important factor determining the performance of the hierarchical controller, as can be seen clearly from the above examples. On the other hand, when $\bar{Q}$ is ill-conditioned, different decompositions may cause largely different $\cond(\hat{Q})$ and  $\cond(\mathcal{P})$. In that case, $\cond(\hat{Q})$ becomes a more important factor determining the performance of the hierarchical controller. An example on this will be shown in Section \ref{sec: application}.
	\end{remark}

\section{Application to Multi-Agent Formation Maneuver Control}\label{sec: application}

\subsection{Problem Formulation}\label{subsec formation}	
Consider a group of agents, denoted by the vertex set $\mathcal{V}$, traveling between multiple waypoints in a two-dimensional plane. The set $\mathcal{V}$ is categorized into two subsets, $\mathcal{V}=\mathcal{L}\cup\mathcal{F}$, where $\mathcal{L}$ is the set of leaders tracking specific positions at each waypoint, and $\mathcal{F}$ is the set of followers adjusting their positions according to measured information from other agents. 
\subsubsection{Agent Dynamics}
To make this example more realistic, we consider that the model of  each agent $i$ is perturbed by a bounded time-varying disturbance $d_i$, so that the dynamics can be written as
\begin{equation}
M_i\ddot{q}_i+C_i\dot{q}_i=u_i+d_i, \;\; i=1,...,N,
\end{equation}	
where $q_i\in\mathbb{R}^2$, $\dot{q}_i\in\mathbb{R}^2$, $M_i\succ0$, $C_i\succ0$, $u_i\in\mathbb{R}^2$ and $d_i\in\mathbb{R}^2$ are the position, the velocity, the inertial matrix, the centrifugal term, the control input, and the disturbance input of agent $i$, respectively. The matrices $M_i$ and $C_i$ are considered to be unknown. The disturbance $d_i$ is assumed to be measurable. Similar assumptions have been made in the existing literature of model-free LQR problems with noise, e.g., \cite{Bian16,Jing20}.

\subsubsection{Control Objective}	
The general control objective is to design a state-feedback control law without knowing $M_i$ and $C_i$ for each agent $i$, such that the agents maintain a desired formation shape at each waypoint, achieve a flocking behavior while traveling between each of two neighboring waypoints, and take the minimum control effort for transforming rom one formation shape to the other. 


\subsubsection{Formation Graph and Communication Graph}

We consider the formation graph $\mathcal{G}_f=(\mathcal{E}_f,\mathcal{V})$ and the communication graph $\mathcal{G}_c=(\mathcal{E}_c,\mathcal{V})$ as two different graphs. Graph $\mathcal{G}_f$, together with the target position $h=(h_1^{\top},...,h_N^{\top})^{\top}\in\mathbb{R}^{2N}$, form a framework $(\mathcal{G}_f,h)$ describing the desired formation. Here $h_i$ is the target position of agent $i$ at the next waypoint. The edges determined by $\mathcal{E}_f$ specify the agent pairs with constrained relative positions,  while the edges determined by $\mathcal{E}_c$ specify those agent pairs who communicate with each other. Usually $\mathcal{E}_c$ is determined by the wireless sensing range of each agent. Therefore, staying cohesive guarantees that there are more agent pairs that can communicate with each other.

\subsubsection{LQR Problem Formulation}

Let $x_i=(q_i^{\top}-h_i^{\top}, \dot q_i^{\top})^{\top}$ for $i=1,...,N$. Then the $i^{th}$ agent model can be rewritten as
\begin{equation}\label{mc dynamics}
\dot x_i=A_ix_i+B_iu_i+B_id_i, ~~i=1,...,N
\end{equation}
where $A_i=\begin{pmatrix}
\mathbf{0} & I_2\\
\mathbf{0} & -M_i^{-1}C_i
\end{pmatrix}$, and $B_i=\begin{pmatrix}
0\\M_i^{-1}
\end{pmatrix}$.
Let $S_1=(I_2,\mathbf{0}_{2\times2})$, $S_2=(\mathbf{0}_{2\times2},I_2)$, then $q_i-h_i=S_1x_i$, and $\dot{q}_i=S_2x_i$. In the literature, asymptotic convergence of formation stabilization has been widely studied \cite{Ren08}. However, the performance of the group trajectory during transience is usually not guaranteed. To capture an optimal trajectory of the agents during travel between waypoints, we make the group minimize the following performance index:
\begin{equation}
\begin{split}
J_1&=\int_0^\infty \sum_{(i,j)\in\mathcal{E}_f}||q_i-q_j-(h_i-h_j)||^2+\sum_{i\in\mathcal{L}}||q_i-h_i||^2dt\\
&=\int_0^\infty x^{\top}(L\otimes S_1^{\top}S_1)x+x^{\top}(\Lambda\otimes I_2)xdt
\end{split}
\end{equation}
where $L\in\mathbb{R}^{N\times N}$ is the Laplacian matrix corresponding to the formation graph $\mathcal{G}_f$, $\Lambda=\diag\{\Lambda_1,...,\Lambda_N\}$, $\Lambda_i=1$ if $i\in\mathcal{L}$ and $\Lambda_i=0$ otherwise.

Flocking behavior would require the agents to stay cohesive and have a common velocity. To model this, we additionally define the following performance index:
\vspace{-0.3mm}
\begin{equation}
J_2=\int_0^{\infty}\dot{q}^{\top}(L\otimes I_2)\dot{q}dt=\int_0^{\infty}x^{\top}(L\otimes S_2^{\top}S_2)xdt.
\end{equation}
The overall goal then is to minimize 
\begin{equation}\label{formation PI}
J=J_1+J_2=\int_0^{\infty}\left[x^{\top}((L+\Lambda)\otimes I_4)x+u^{\top}u\right]dt
\end{equation}
subject to the plant dynamics (\ref{mc dynamics}) for $i=1,\dots,N$.

\begin{remark}
	Different from formation control problems, where the steady states of all agents depend on their initial states, in the maneuver control problem each agent has a specific target position at each waypoint. In this case, it is easy for each agent to obtain its own as well as other's target steady states by either (i) using available target relative positions from its neighbors and communicating with them, or (ii) achieving the target position through a centralized task assignment.
\end{remark}

	\subsection{Feasible Decompositions for Hierarchical Control}\label{subsec formation decomposition}
	
	To use our hierarchical control design, Assumptions \ref{as control and observe} and \ref{as Qj observable} should be satisfied. Let $\mathcal{A}=\diag\{A_1,...,A_N\}$ and $\mathcal{B}=\diag\{B_1,...,B_N\}$, where $A_i$ and $B_i$ are shown in (\ref{mc dynamics}). Since $M_i\succ0$ and $C_i\succ0$ for each agent $i$, $(\mathcal{A},\,\mathcal{B})$ must be controllable. Moreover, when graph $\mathcal{G}_f$ is connected and there exists at least one leader in the whole group, it  holds that $Q=(L+\Lambda)\otimes I_4\succ0$, implying that $(Q^{1/2},\,\mathcal{A})$ is observable. Then Assumption \ref{as control and observe} is satisfied. To ensure validity of Assumption \ref{as Qj observable}, the decomposition must satisfy the following result.
	

	\begin{theorem}\label{th formation leader}
	 Assumption \ref{as Qj observable} is satisfied if and only if each cluster contains at least one leader and its formation graph is connected.\QEDA
	\end{theorem}
	
An immediate consequence of Theorem \ref{th formation leader} is that the number of clusters cannot exceed the number of leaders in the MAS.
To find the decomposition maximizing $\kappa$, we modify the MIQP (\ref{MIQP}) to include the following two constraints: 

(i) $\xi^{\top}\eta_i\geq1$ for $i=1,...,s$, where let  $\xi\in\mathbb{R}^N$ with $\xi(i)=1$ if $i\in\mathcal{L}$ and $\xi(i)=0$ otherwise; 

(ii) $G_{kk}-\sum_{i=1}^s\eta_i^{\top}g_ke_k^{\top}\eta_i\geq\epsilon(\sum_{i=1}^s\eta_i^{\top}\mathbf{1}_Ne_k^{\top}\eta_i-1)/N$, $k=1,...,N$, where $g_k\in\mathbb{R}^N$ is the $k$-th column of $G$, and $\epsilon$ is the minimum nonzero entry of $\bar{G}$. 

The constraint (i) implies that each cluster contains at least one leader, while constraint (ii) means that each agent $k$ either forms a cluster ($\sum_{i=1}^s\eta_i^{\top}\mathbf{1}_Ne_k^{\top}\eta_i=1$) or has a nonempty neighbor set in the cluster that it belongs to.

\vspace{-3mm}
\subsection{Numerical Experiments}

\begin{figure}[t]
	\centering
	\includegraphics[width=0.5\textwidth]{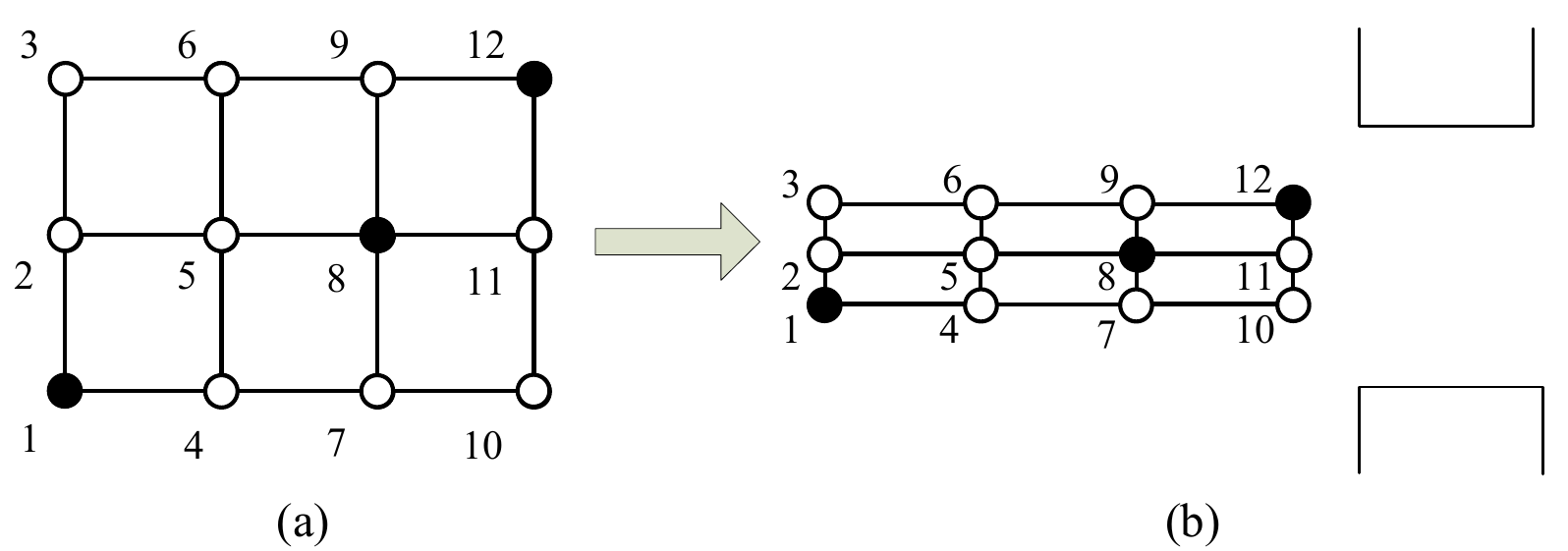}
	\caption{(a). The initial positions of agents; (b) The target positions of agents at the next waypoint.}
	\label{fig transformation}
\end{figure}

We consider 12 agents governed by (\ref{mc dynamics}) with the formation graph $\mathcal{G}_f$ as a mesh grid shown in Fig. \ref{fig transformation}. The black nodes are leaders, and the  other nodes are followers. The mission of these agents is to transform their formation shape from Fig. \ref{fig transformation} (a) to Fig. \ref{fig transformation} (b) before entering into a narrow space as shown towards the right in the figure. The leader set is $\mathcal{L}=\{1,8,12\}$. The parameters in the dynamics of each agent $i$ are set as $M_i=\diag\{(i+1)/2,i/2\}$ and $C_i=\diag\{i/4,i/5\}$. The disturbance is set as $d_i(t)=\left((0.05i, 0.1i)^{\top}\times\cos t\right)/(t+1)$ for $i=1,...,N$. 

\begin{figure}
	\centering
	\includegraphics[width=0.5\textwidth]{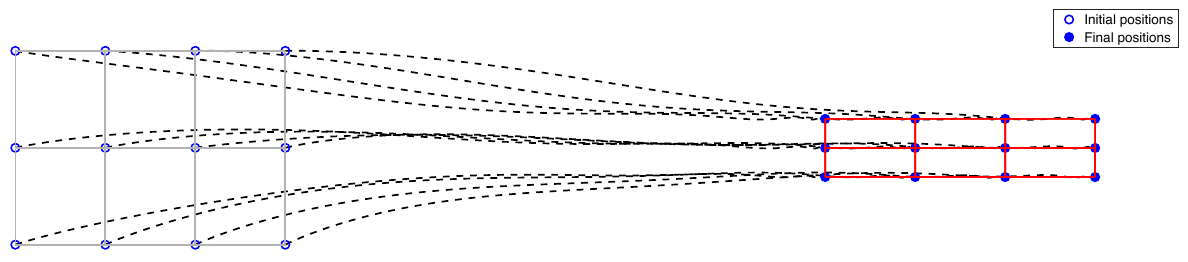}
	\caption{The optimal trajectory for maneuver control.}
	\label{fig undecompose}
\end{figure}

\begin{figure}[t]
	\centering
	\includegraphics[width=0.5\textwidth]{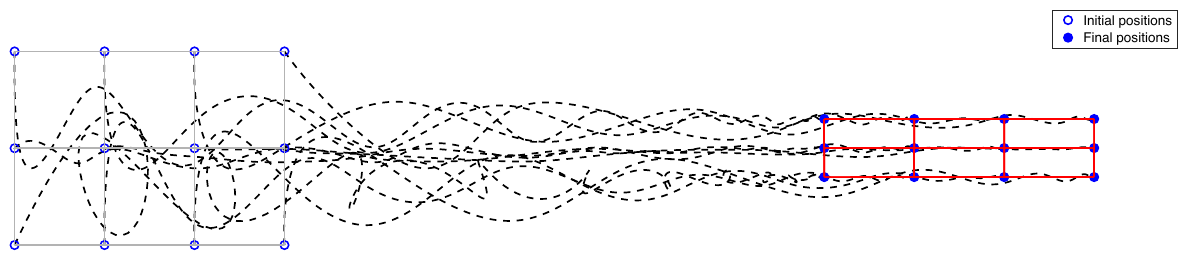}
	\caption{State trajectories for maneuver control using the formation stabilization law from \cite{Ren08}.}
	\label{fig stabilization}
\end{figure}

We first find the optimal LQR without considering the disturbance $d_i$. By implementing the optimal control law in the presence of disturbance, the value of the performance index is found as $J^*=1247.7575$, and that for the control effort as $J_u=\int_0^\infty u^{\top}udt=428.8475$. The number of communication links is 66, which implies that the communication graph $\mathcal{G}_c=(\mathcal{V},\mathcal{E}_c)$ is fully connected. Under the same communication graph $\mathcal{G}_c$, by using the formation stabilization law in \cite{Ren08} we get
\begin{equation}\label{stabilization law}
u_i=-\sum_{(i,j)\in\mathcal{E}_c}\left(q_i-q_j-(h_i-h_j)\right)-k_i(q_i-h_i)
\end{equation}
where $k_i=1$ if $i\in\mathcal{L}$ and $k=0$ otherwise. The overall performance and control input performance are computed as $J=2176.9360$ and $J_u=983.9627$, respectively. Fig. \ref{fig undecompose} and Fig. \ref{fig stabilization} show the state trajectories of the agents corresponding to the optimal LQR and the formation stabilization law (\ref{stabilization law}), respectively. From these two figures we observe that compared to the stabilization law (\ref{stabilization law}), LQR provides a better closed-loop path response and thereby saves the overall control effort. 

Next, we compute the model-free LQR using the conventional ADP algorithm from \cite{Jiang12}. In the data collection phase, $u_i+d_i$ (instead of $u_i$) is viewed as the control input. By implementing the learned controller, the overall performance and control input performance are computed as $J^{RL}=1247.7601$ and $J_u^{RL}=429.2707$, respectively. The learning time is 58.8262s.

\begin{figure}
	\centering
	\includegraphics[width=0.5\textwidth]{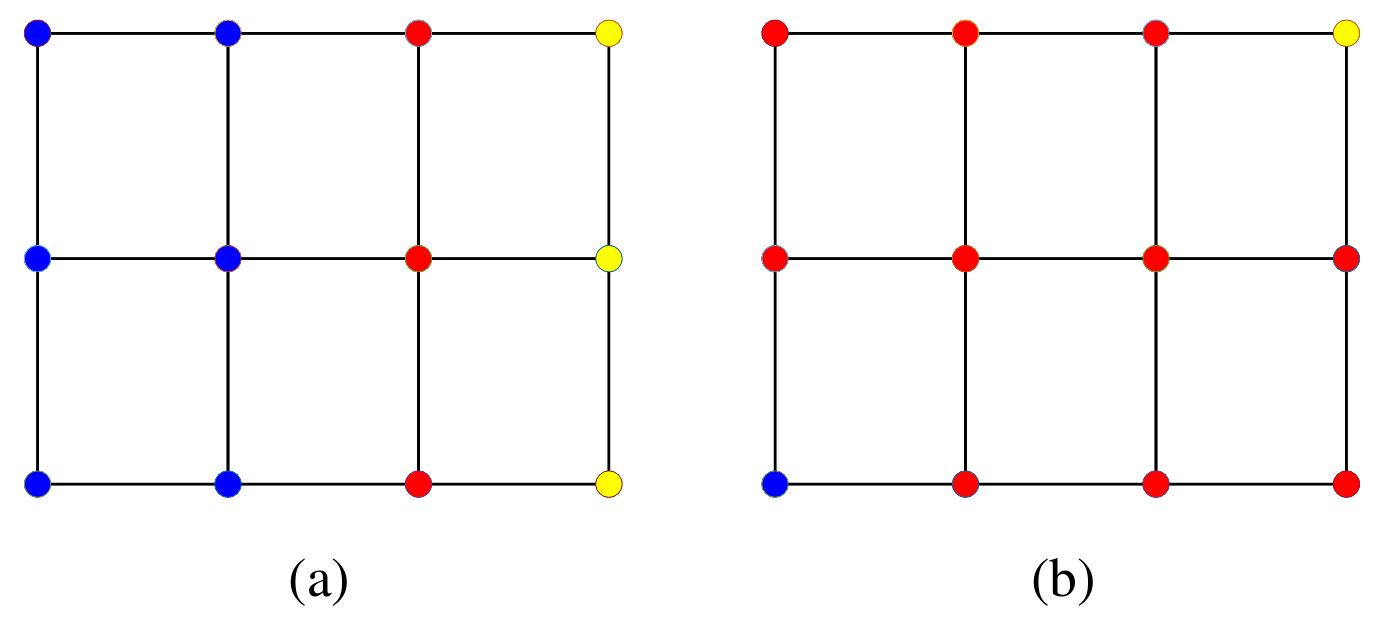}
	\caption{(a). The decomposition maximizing $\kappa$; (b) The decomposition minimizing $\tr(G_2)$.}
	\label{fig decomposition}
\end{figure}

\begin{figure}[t]
	\centering
	\includegraphics[width=0.5\textwidth]{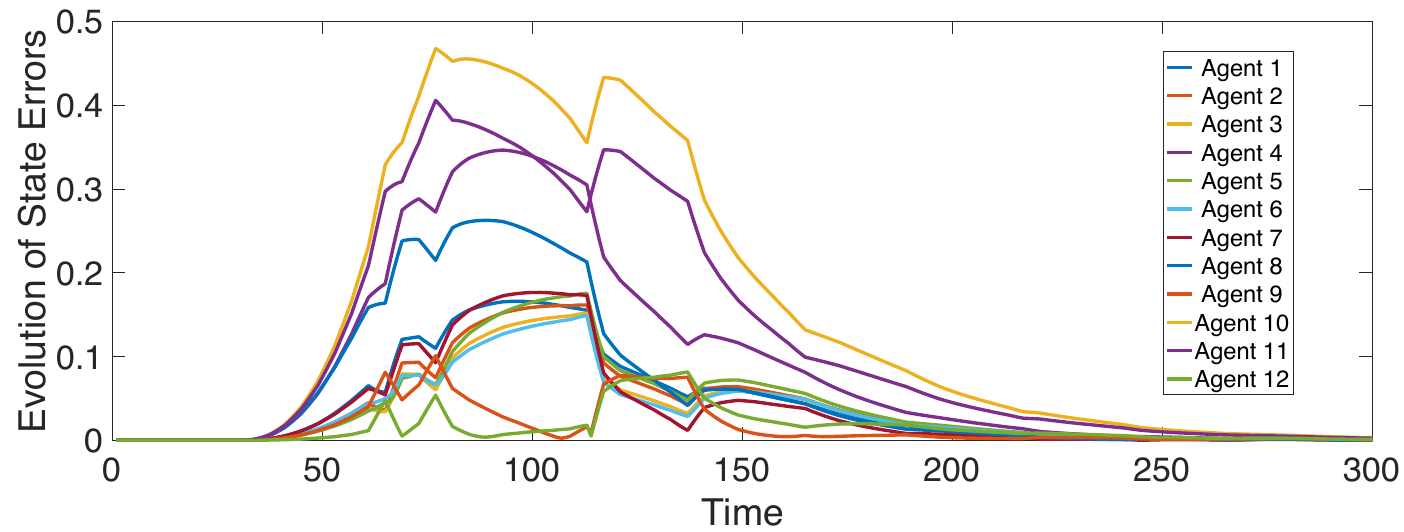}
	\caption{The evolution of agent state errors by implementing the hierarchical controller with the decomposition maximizing $\kappa$ and the optimal LQR, respectively.}
	\label{fig stateerror1}
	\vspace{-0.6cm}
\end{figure}

\begin{figure}[t]
	\centering
	\includegraphics[width=0.5\textwidth]{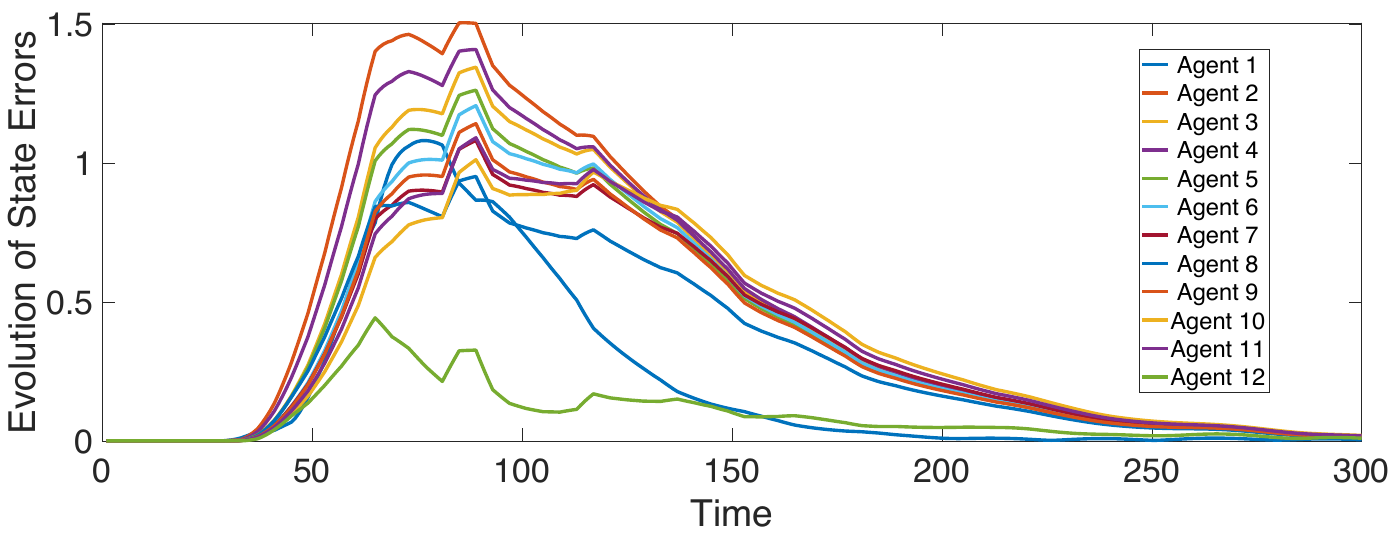}
	\caption{The evolution of agent state errors by implementing the hierarchical controller with the decomposition minimizing $\tr(G_2)$ and the optimal LQR, respectively.}
	\label{fig stateerror2}
\end{figure}

Finally, we apply our proposed hierarchical approximation. Similar to when applying conventional RL, we regard $u_i+d_i$ as the control input during the data collection phase. For validity of Assumption \ref{as Qj observable}, according to Theorem \ref{th formation leader}, we decompose the group into $s=3$ clusters such that each cluster has a connected formation graph, and contains at least one leader. In Fig. \ref{fig decomposition}, the decomposition strategies maximizing $\kappa$ and minimizing $\tr(G_2)$ are shown, respectively, where the nodes with the same color belong to the same cluster. The first decomposition is obtained by solving (\ref{MIQP}) with the additional constraints that were presented in Subsection \ref{subsec formation decomposition}. The learning time is found to be 22.7734 seconds. Fig. \ref{fig stateerror1} shows the evolution of the position error of the agents by comparing the hierarchical control law with the decomposition maximizing $\kappa$ and the optimal LQR. The second decomposition is obtained by solving a MILP formulation for the minimum $s$-cut problem with additional constraints in Subsection \ref{subsec formation decomposition}. Fig. \ref{fig stateerror2} shows the evolution of the position errors by comparing the hierarchical control law with the decomposition minimizing $\tr(G_2)$ and the optimal LQR. From Fig. \ref{fig stateerror1} and Fig. \ref{fig stateerror2}, we observe that the decomposition maximizing $\kappa$ has a better performance than the decomposition minimizing $\tr(G_2)$.
		
	More detailed simulation results of implementing the hierarchical RL algorithm (Algorithm \ref{alg:2}) based on three decompositions are listed in Table \ref{tab1}. For simplicity, each decomposition is done according to the order of the agents. As an example, for the first decomposition in Table \ref{tab1}, the indices of agents in the three clusters are $\{1,...,6\}$, $\{7,8,9\}$ and $\{10,11,12\}$. It is observed from Table \ref{tab1} that the first decomposition maximizes $\kappa$, and induces the fewest communication links required by the hierarchical controller. The second decomposition minimizes $\tr(G_2)$, but does not induce the best performance. Instead, the first decomposition, which induces the minimum condition numbers on  $\mathcal{P}$ and $\hat{Q}$, yields the best performance. This is because $\cond(\bar{Q})=\cond(\Lambda)=\infty$, which implies that $\bar{Q}$ is ill-conditioned. In this case, different decomposition strategies lead to largely different $\cond(\mathcal{P})$ and $\cond(\hat{Q})$, and thus $\tr(G_2)$ becomes a less important factor to determine the suboptimality. 
	
	\begin{remark}
	The inter-agent interactions during implementation of Algorithm \ref{alg:2} can be explained as follows. Suppose we choose the first decomposition strategy in Table \ref{tab1}. During the implementation of Algorithm 2, no communication is required between agents $\{1, ..., 6\}$ and agents $\{10,11,12\}$. The communication between the rest of agents can be achieved through three coordinators in the three clusters. In other words, the hierarchical RL algorithm can be implemented distributively, as described in Remark \ref{re distributed learning}.
	\end{remark}
	
	
\begin{table*}[htbp]	
	\centering
	\fontsize{8}{9}\selectfont
	\begin{threeparttable}
		\caption{Comparisons between different decompositions.}
		\label{tab1}
		\begin{tabular}{cccccccccccc}
			\toprule			
			\multicolumn{3}{c}{Decomposition}&\multicolumn{9}{c}{Performance Indices}\cr
			\cmidrule(lr){1-3} \cmidrule(lr){4-12}
			$N_1$&$N_2$&$N_3$&$\kappa$&$\tr(G_2)$&$\cond(\mathcal{P})$&$\cond(\hat{Q})$&$J$&$J_u$&$n_c$&Time(sec)&SOP\cr
			\midrule
			6&3& 3 & 18 & 12&248.7647&46.1346& 1259.7985&426.9677&48&0.9248&0.82\%\cr			
			1&10&1& 1 & 8&339.4430&83.7524& 1347.0390&431.6374&65&13.9719&7.96\%\cr
			7&2& 3  & 0& 12& 285.1161 & 55.3510& 1267.7974&442.4165&66&2.1165&1.61\%\cr
			\bottomrule
		\end{tabular}
	\end{threeparttable}
\end{table*}

\section{Conclusion}\label{sec: conclusion}
We presented a model-free hierarchical RL algorithm for optimal control of linear MAS with heterogeneous agents, based on a decomposition approach that can significantly reduce learning time. The derived controller is suboptimal but has a specific structure. Two indices of this hierarchical controller, namely, the number of communication links and the closed-loop performance of the agents, are analyzed and shown to be dependent on the decomposition strategy. Optimizing  these two indices is formulated as a MIQP and a minimum $s$-cut problem, respectively. The hierarchical controller is applied to a formation maneuver control problem. Simulation results are presented to illustrate its effectiveness in saving learning time, and corresponding trade-offs in closed-loop performance.

%

\vspace{-4mm}

\section{Appendix}\label{appendix}

{\it Proof of Theorem \ref{th stability}:}
Because $\mathcal{P}$ satisfies~\eqref{locric} with Assumption~\ref{as control and observe} and~\ref{as Qj observable}, $\mathcal{P}$ is positive definite.  Consider the Lyapunov function $\Phi=x^\top \mathcal{P} x$, whose derivative is given by
\begin{align}
\dot \Phi &= 2x^\top \mathcal{P}(\mathcal{A}x + \mathcal{B}u_h) \\
&=2x^\top \mathcal{P}(\mathcal{A}-\mathcal{B}{R^{-1}\mathcal{B}^\top\mathcal{P}}-\mathcal{B}\tilde{R}\mathcal{B}^\top\mathcal{P})x\\
&=2x^\top (\mathcal{P}\mathcal{A}-\mathcal{P}\mathcal{B}{R^{-1}\mathcal{B}^\top\mathcal{P}})x-2x^\top \mathcal{P}\mathcal{B}\tilde{R}\mathcal{B}^\top\mathcal{P}x\\
&=x^\top (\mathcal{P}\mathcal{A}+\mathcal{A}^\top\mathcal{P}-2\mathcal{P}\mathcal{B}{R^{-1}\mathcal{B}^\top\mathcal{P}})x-2x^\top \mathcal{P}\mathcal{B}\tilde{R}\mathcal{B}^\top\mathcal{P}x\\
&=x^\top (-\hat Q-\mathcal{P}\mathcal{B}{R^{-1}\mathcal{B}^\top\mathcal{P}})x-2x^\top \mathcal{P}\mathcal{B}\tilde{R}\mathcal{B}^\top\mathcal{P}x\leq 0. 
\end{align}
From LaSalle's Invariance Principle, we know that $x$ converges to the largest invariant set $\mathcal{E}$ defined as 
\begin{equation}
\mathcal{E}=\{x~|~x^\top \hat Qx=x^\top\mathcal{P}\mathcal{B}{R^{-1}\mathcal{B}^\top\mathcal{P}}x=2x^\top \mathcal{P}\mathcal{B}\tilde{R}\mathcal{B}^\top\mathcal{P}x=0\}.
\end{equation}
Since $\hat Q$ is positive semidefinite, it follows that $\hat Q^{1/2}x = 0$ in $\mathcal{E}$. From (\ref{hierarchcial controller}), $u_h=0$ in $\mathcal{E}$. Thus, we consider the dynamics in $\mathcal{E}$ as $\dot x = \mathcal{A}x$, and $\hat Q^{1/2}x = 0$. It follows that $\hat Q^{1/2}\mathcal{A}^cx=0$ for $c\in\{0,1,2,...\}$. Since $(\hat Q^{1/2}, \mathcal{A})$ is observable, we have $x=0$. Then we conclude that the only solution that stays in $\mathcal{E}$ is the trivial solution $x=0$ and, therefore, the closed-loop system is globally asymptotically stable. 
\QEDA

{\it Proof of Theorem \ref{th structure}: }
From the uniqueness of Moore-Penrose inverse, we can define $\Xi=\diag\{\Xi_1,...,\Xi_s\}\in\mathbb{R}^{mN\times nN}$, where $\Xi_j=(\mathcal{P}_j\mathcal{B}_j)^+\in\mathbb{R}^{m_j\times n_j}$. Let $\Phi=G_2\otimes\tilde{Q}$. Next, we partition $\Phi$ into $s^2$ blocks $\Phi_{ij}\in\mathbb{R}^{n_i\times n_j}$, $i,j=1,...,s$, according to the $s$ clusters. For two different $i$ and $j$ such that $G_2(i,j)=\mathbf{0}_{N_i\times N_j}$, we have $\Phi_{ij}=\mathbf{0}_{n_i\times n_j}$. As a result, 
\begin{equation}
\tilde{R}(i,j)=\Xi_i\Phi_{ij}\Xi_j^{\top}=\mathbf{0}_{m_i\times m_j}.
\end{equation}
From (\ref{hierarchcial controller}), the hierarchical controller of the $j$-th cluster is in the following form
\begin{equation}
\mathbf{u}_j=-\hat{R}_j^{-1}\mathcal{B}_j^{\top}\mathcal{P}_j\mathbf{x}_j-\sum_{i=1}^s\tilde{R}(i,j)\mathcal{B}_i^{\top}\mathcal{P}_i\mathbf{x}_i,
\end{equation}
which implies that communication between cluster $i$ and cluster $j$ is not required if $\tilde{R}(i,j)=\mathbf{0}_{m_i\times m_i}$.
\QEDA

{\it Proof of Theorem \ref{th JJJ}: }
The second inequality $ J(x(0),u^*)\leq J(x(0),u_h^*)$ is valid because $u^*$ is the optimal controller corresponding to $J(x(0),u)$. Next, we prove the first inequality.

From (\ref{locric}), we observe that $\mathcal{P}$ is also the solution to
\begin{equation}\label{P hatQ}
\mathcal{P}\mathcal{A}+\mathcal{A}^{\top}\mathcal{P}+\hat{Q}-\mathcal{P}\mathcal{B}R^{-1}\mathcal{B}^{\top}\mathcal{P}=0.
\end{equation}
Note that $Q-\hat{Q}=G_2\otimes\tilde{Q}\succeq0$, i.e., $\hat{Q}\leq Q$. Using (\ref{original P}) and \cite[Lemma 3]{Willems71}, we obtain $\mathcal{P}\leq P$. It follows that $\mathcal{J}(x(0),u_h^*)\leq J(x(0),u^*)$.
\QEDA

{\it Proof of Theorem \ref{th performance gap}: }
We first study $\mathbb{E}(J(x(0),u_h^*))$. We can write
\begin{equation}
\begin{split}
J(x(0),u_h^*)&=\int_0^\infty x^{\top}Qxdt+\int_0^\infty{u^*_h}^{\top}Ru_h^*dt\\
&=\int_0^\infty x^{\top}(Q+K_h^{\top}RK_h)xdt
\end{split}
\end{equation}
where $x=e^{(\mathcal{A}-\mathcal{B}K_h)t}x(0)$. It follows that 
\begin{equation}
\mathbb{E}( J(x(0),u_h^*))=\tr(U\mathbb{E}(x(0)x^{\top}(0)))=\sigma^2\tr(U),
\end{equation}
where $U$ is the solution of 
\begin{equation}\label{U equation}
\mathcal{A}^{\top}_sU+U\mathcal{A}_s+Q+K_h^{\top}RK_h=0,
\end{equation}
with $\mathcal{A}_s=\mathcal{A}-\mathcal{B}K_h$.

Next, we analyze $\mathbb{E}(J(x(0),u^*))$. It can be verified that 
$$\mathbb{E}(J(x(0),u^*))=\mathbb{E}(x^{\top}(0)Px(0))=\sigma^2\tr(P),$$ 
where $P$ is the solution of (\ref{original P}). Let $V=U-P$. Then $\mathbb{E}(\Delta J_h)=\mathbb{E}(J(x(0),u_h^*))-\mathbb{E}(J(x(0),u^*))=\sigma^2\tr(V)$. 
Note that (\ref{original P}) can be rewritten as
\begin{equation}\label{P equation}
P\mathcal{A}_s+\mathcal{A}_s^{\top}P+Q+K^{\top}RK_h+K_h^{\top}RK-K^{\top}RK=0.
\end{equation}
The subtraction of (\ref{U equation}) and (\ref{P equation}) yields (\ref{V equation}).
\QEDA

{\it Proof of Lemma \ref{le bound for trace}:}		
(i) Using (\ref{locric}), the positive definiteness of $\hat{Q}$, and \cite[Corollary 4.5.11]{Horn12}, the following holds:
\begin{align*}
&-\lambda_{\max}(\mathcal{P}\mathcal{A}_S\mathcal{P}^{-1}+\mathcal{A}_S^{\top})\\
&=-\lambda_{\max}(\mathcal{P}\mathcal{A}\mathcal{P}^{-1}-\mathcal{P}\mathcal{B}\mathcal{R}^{-1}\mathcal{B}^{\top}+\mathcal{A}^{\top}-\mathcal{P}\mathcal{B}\mathcal{R}^{-1}\mathcal{B}^{\top})\\
&=-\lambda_{\max}(-\mathcal{P}\mathcal{B}\mathcal{R}^{-1}\mathcal{B}^{\top}-\mathcal{Q}\mathcal{P}^{-1} )\\
&=-\lambda_{\max}(-\mathcal{P}^{1/2}\mathcal{B}\mathcal{R}^{-1}\mathcal{B}^{\top}\mathcal{P}^{1/2}-\mathcal{P}^{-1/2}\mathcal{Q}\mathcal{P}^{-1/2})\\
&=\lambda_{\min}(\mathcal{P}^{1/2}\mathcal{B}\mathcal{R}^{-1}\mathcal{B}^{\top}\mathcal{P}^{1/2}+\mathcal{P}^{-1/2}\mathcal{Q}\mathcal{P}^{-1/2})\\
&=\lambda_{\min}(\mathcal{P}^{1/2}\mathcal{B}R^{-1}\mathcal{B}^{\top}\mathcal{P}^{1/2}+\mathcal{P}^{-1/2}\bar{Q}\mathcal{P}^{-1/2})\\
&\geq\lambda_{\min}(\mathcal{P}^{1/2}\mathcal{B}R^{-1}\mathcal{B}^{\top}\mathcal{P}^{1/2})+\lambda_{\min} (\mathcal{P}^{-1/2}\bar{Q}\mathcal{P}^{-1/2})\\
&\geq\lambda_{\min}(\mathcal{P})\lambda_{\min}(\mathcal{B}R^{-1}\mathcal{B}^{\top}) +\lambda_{\min}(\bar{Q})/\lambda_{\max}(\mathcal{P})>0.
\end{align*}
In most applications, usually $\mathcal{B}$ is not square, implying that $\lambda_{\min}(\mathcal{B}R^{-1}\mathcal{B}^{\top})=0$. Therefore, we further omit the term associated with $\lambda_{\min}(\mathcal{B}R^{-1}\mathcal{B}^{\top})$, i.e., 
\begin{equation}\label{muAS}
-\lambda_{\max}(\mathcal{P}\mathcal{A}_S\mathcal{P}^{-1}+\mathcal{A}_S^{\top})\geq\lambda_{\min}(\bar{Q})/\lambda_{\max}(\mathcal{P}).
\end{equation}
From \cite[Corollary 3.2]{Fang97}, it holds that:
\begin{equation}\label{trV}
\tr(V)\leq-\frac{\lambda_{\max}(\mathcal{P})\tr(\mathcal{P}^{-1}W)}{\lambda_{\max}(\mathcal{P}\mathcal{A}_S\mathcal{P}^{-1}+\mathcal{A}_S^{\top})}.
\end{equation}
Reusing \cite[Corollary 4.5.11]{Horn12}, we have
\begin{equation}\label{trP-1W}
\begin{split}
\tr(\mathcal{P}^{-1}W)&=\tr(\mathcal{P}^{-1/2}W\mathcal{P}^{-1/2})\\
&\leq\lambda_{\max}^2(\mathcal{P}^{-1/2})\tr(W)
=\frac{\tr(W)}{\lambda_{\min}(\mathcal{P})}
\end{split}
\end{equation}
Combining (\ref{muAS}), (\ref{trV}) and (\ref{trP-1W}), the bound on $\tr(U)$ stated in (\ref{bound of traceU}) is obtained.

(ii) It can be verified that 
\begin{equation}
W=\mathcal{P}\tilde{M}\mathcal{P}+\mathcal{P}\bar{M}\mathcal{P}+\Delta P\mathcal{M}\Delta P-P\tilde{M}P,
\end{equation}
where $\mathcal{M}=\mathcal{B}\mathcal{R}^{-1}\mathcal{B}^{\top}$, $\tilde{M}=\mathcal{B}\tilde{R}\mathcal{B}^{\top}$, $\bar{M}=\mathcal{B}\tilde{R}R\tilde{R}\mathcal{B}^{\top}$, and $\Delta P=P-\mathcal{P}$. We have shown in Theorem \ref{th JJJ} that $\mathcal{P}\leq P$. It follows that $\tr(P^2-\mathcal{P}^2)=\tr\left((P+\mathcal{P})^{1/2}(P-\mathcal{P})(P+\mathcal{P})^{1/2}\right)\geq0$. As a result, $\tr(\mathcal{P}\tilde{M}\mathcal{P}-P\tilde{M}P)=\tr((\mathcal{P}^2-P^2)\tilde{M})=-\tr\left(\tilde{M}^{1/2}(P-\mathcal{P})\tilde{M}^{1/2}\right)\leq-\lambda_{\min}(\tilde{M})\tr(P^2-\mathcal{P}^2)\leq0$. This implies that $\tr(W)\leq\tr(\mathcal{P}\bar{M}\mathcal{P})+\tr(\Delta P\mathcal{M}\Delta P)$. 
To prove the statement (ii), we will prove $\tr(\mathcal{P}\bar{M}\mathcal{P})\leq f_1$ and $\tr(\Delta P\mathcal{M}\Delta P)\leq f_2$ successively.

For the first inequality, by \cite[Corollary 4.5.11]{Horn12}, the following holds:
\begin{equation}\label{f1 deriv}
\begin{split}
&\tr(\mathcal{P}\mathcal{B}\tilde{R}R\tilde{R}\mathcal{B}^{\top}\mathcal{P})\leq \lambda_{\max}^2((\mathcal{P}\mathcal{B})(\mathcal{P}\mathcal{B})^+)\times\\
&\quad\tr\left((G_2\otimes\tilde{Q})(\mathcal{P}\mathcal{B})^{+T}R(\mathcal{P}\mathcal{B})^+(G_2\otimes\tilde{Q})\right)\\
&=\tr\left (R^{1/2}(\mathcal{P}\mathcal{B})^+(G_2\otimes\tilde{Q})^2(\mathcal{P}\mathcal{B})^{+T}R^{1/2}\right)\\
&\leq\tr^2(G_2\otimes\tilde{Q})\lambda_{\max}\left[(\mathcal{P}\mathcal{B})^{+T}R(\mathcal{P}\mathcal{B})^+\right],\\
&=\tr^2(G_2)\tr^2(\tilde{Q})\lambda_{\max}\left(SRS\right),
\end{split}
\end{equation}
where $S^2 = (\mathcal{P}\mathcal{B})^+(\mathcal{P}\mathcal{B})^{+T}$, i.e., $S=T\Lambda^{1/2}T^\top$ in which $T$ and $\Lambda$ follow from the eigen-decomposition of $ (\mathcal{P}\mathcal{B})^+(\mathcal{P}\mathcal{B})^{+T}$ satisfying $T\Lambda T^\top= (\mathcal{P}\mathcal{B})^+(\mathcal{P}\mathcal{B})^{+T}$. Since $S$ is square and symmetric, we can apply~\cite[Corollary 4.5.11]{Horn12} to analyze the eigenvalues of $SRS$. 

The singular values of $S$ are given by the diagonal entries of $\Lambda^{1/2}$. The minimum singular value is zero while the maximum singular value is less than $\sigma_{\max}((\mathcal{P}\mathcal{B})^+)$. Therefore, we have
\begin{equation}
\begin{split}
\lambda_{\max}(SRS)\leq \sigma^2_{\max}((\mathcal{P}\mathcal{B})^+)\lambda_{\max}(R).
\end{split}
\end{equation}
Using the fact that  $\sigma_{\max}((\mathcal{P}\mathcal{B})^+)=1/\sigma_{l}(\mathcal{P}\mathcal{B})$, where $\sigma_l(\mathcal{P}\mathcal{B})$ is the minimum nonzero singular value of $\mathcal{P}\mathcal{B}$, we next look for a lower bound on $\sigma_l(\mathcal{P}\mathcal{B})$.

Note that $\sigma_l(\mathcal{P}\mathcal{B})=\sigma_l(\mathcal{B}^{\top}\mathcal{P}).$
From \cite[Corollary 4.5.11]{Horn12}, it holds that 
\begin{equation}
\begin{split}
\sigma_l(\mathcal{B}^{\top}\mathcal{P})&=\lambda_l^{1/2}(\mathcal{P}\mathcal{B}\mathcal{B}^{\top}\mathcal{P})\\&\in[\lambda_{\min}(\mathcal{P})\sigma_l(\mathcal{B}), \lambda_{\max}(\mathcal{P})\sigma_{\max}(\mathcal{B})].
\end{split}
\end{equation}
We then have
\begin{equation}
\begin{split}
\lambda_{\max}(SRS)&\leq\frac{1}{ \lambda^2_{\min}(\mathcal{P})\sigma^2_l(\mathcal{B})}\lambda_{\max}(R).
\end{split}
\end{equation}	
Together with (\ref{f1 deriv}), we have $\tr(\mathcal{P}\bar{M}\mathcal{P})\leq f_1$.

In the following derivations for $f_2$, some steps that are similar to the approach stated above will be omitted for brevity. We use the following two inequalities:
\begin{equation}
\begin{split}
\tr(\Delta P\mathcal{B}R^{-1}\mathcal{B}^{\top}\Delta P)\leq \lambda_{\max}&(\Delta P)\tr(\mathcal{B}\tilde{R}\mathcal{B}^{\top})\\
\leq (\lambda_{\max}&(P)-\lambda_{\min}(\mathcal{P}))\tr(\mathcal{B}\tilde{R}\mathcal{B}^{\top}),
\end{split}		
\end{equation}
\begin{equation}
\begin{split}
\tr(\Delta P\mathcal{B}\tilde{R}\mathcal{B}^{\top}\Delta P)\leq \lambda_{\max}&(\Delta P)\sigma_{\max}^2(\mathcal{B})\tr(\tilde{R})\\
\leq\left[\lambda_{\max}(P)-\lambda_{\min}(\mathcal{P})\right]&\sigma_{\max}^2(\mathcal{B})\frac{\tr(G_2\otimes\tilde{Q})}{\lambda_{\min}^2(\mathcal{P})\sigma_l^2(\mathcal{B})}.
\end{split}
\end{equation}
Combing the two inequalities, we get $\tr(\Delta P\mathcal{M}\Delta P)\leq f_2$.
\QEDA

{\it Proof of Theorem \ref{th simple case}:} 
Without loss of generality, given $s$ as the number of clusters, and the decomposition $E=(\eta_1,...,\eta_s)$ corresponding to the maximum $\kappa=\kappa^*(s)=\sum_{i\nsim j}N_iN_j$. Next we show that for $s+1$, there exists a decomposition $\bar{E}=(\eta_1,...,\eta_{s+1})$ such that the corresponding $\kappa'\geq\kappa^*(s)$. For any cluster with at least two agents, if we decompose it into two clusters $k$ and $l$, the resulting $\kappa$ will be $\kappa'=\kappa^*(s)+N_kN_l>\kappa^*(s)$ if $k\nsim l$, and $\kappa'=\kappa^*(s)$ otherwise. This shows the nondecreasing property of $\kappa^*(s)$. Since $s\leq N$, and $\kappa(N)=z_0$, we conclude that $\kappa$ is bounded by $z_0$.
\QEDA

{\it Proof of Theorem \ref{th formation leader}: }
For each cluster $j$, the performance index can be written as $J=\int_0^\infty [\mathbf{x}_j^{\top}((L_j+\Lambda_j)\otimes I_2)+u^{\top}u]dt$, where $L_j$ is the Laplacian matrix corresponding to the subgraph $\mathcal{G}_f^j$ of $\mathcal{G}_f$ involving agents in $\mathcal{V}_j$, $\Lambda_j$ is a diagonal matrix.

Sufficiency: When there is a leader in cluster $j$, there is at least one positive value on diagonal of $\Lambda_j$. Since $\mathcal{G}_f^j$ is connected, using \cite[Lemma 3]{Hong06}, we have $\hat{Q}_j=(L_j+\Lambda_j)\otimes I_4\succ0$. As a result, $(\hat{Q}_j^{1/2},\mathcal{A}_j)$ is observable.

Necessity: Suppose that no leaders exist in cluster $j$ or $\mathcal{G}_f^j$ is not connected. Then $\hat{Q}_j=L_j\otimes I_4$ and $L_j$ has at least one zero eigenvalue. Let $\zeta$ be the eigenvector associated with eigenvalue 0 of $L_j$. We observe that $\hat{Q}_j$ and $\mathcal{A}_j$ have a common eigenvector corresponding to eigenvalue 0, which is $\zeta\otimes (1, 1, 0, 0)^{\top}$. Therefore, $(\hat{Q}_j^{1/2},\mathcal{A}_j)$ can never be detectable.
\QEDA	

\vspace{-0.5cm}
\begin{IEEEbiography}[{\includegraphics[width=1in,height=1.25in,clip,keepaspectratio]{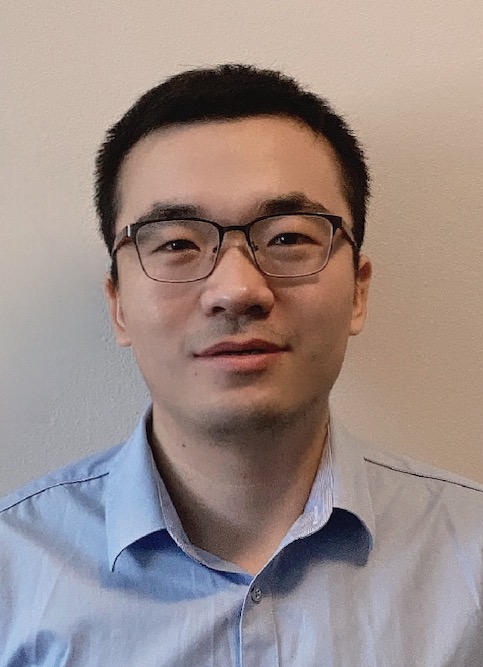}}]{Gangshan Jing} received the Ph.D. degree in Control Theory and Control Engineering from Xidian University, Xi'an, China, in 2018. He has been a postdoctoral researcher in Department of Electrical and Computer Engineering, at North Carolina State University, USA, since 2019 Sept.. He was a research assistant and a postdoctoral researcher at Hong Kong Polytechnic University and Ohio State University in 2016-2017 and 2018-2019, respectively. His research interests include control, optimization, and machine learning for network systems.
\end{IEEEbiography} 
\vspace{-0.5cm}
\begin{IEEEbiography}[{\includegraphics[width=1in,height=1.25in,clip,keepaspectratio]{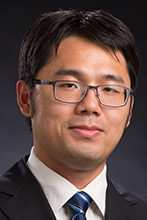}}]{He Bai} received his Ph.D. degree in Electrical Engineering from Rensselaer Polytechnic Institute, Troy, NY, in 2009. From 2009 to 2010, he was a postdoctoral researcher at Northwestern University, Evanston, IL. From 2010 to 2015, he was a Senior Research and Development Scientist at UtopiaCompression Corporation, Los Angeles, CA. In 2015, he joined the School of Mechanical and Aerospace Engineering at Oklahoma State University, Stillwater, OK, as an assistant professor. His research interests include distributed estimation, control and learning, reinforcement learning,  nonlinear control, and robotics. 
\end{IEEEbiography} 
\vspace{-0.5cm}
\begin{IEEEbiography}[{\includegraphics[width=1in,height=1.25in,clip,keepaspectratio]{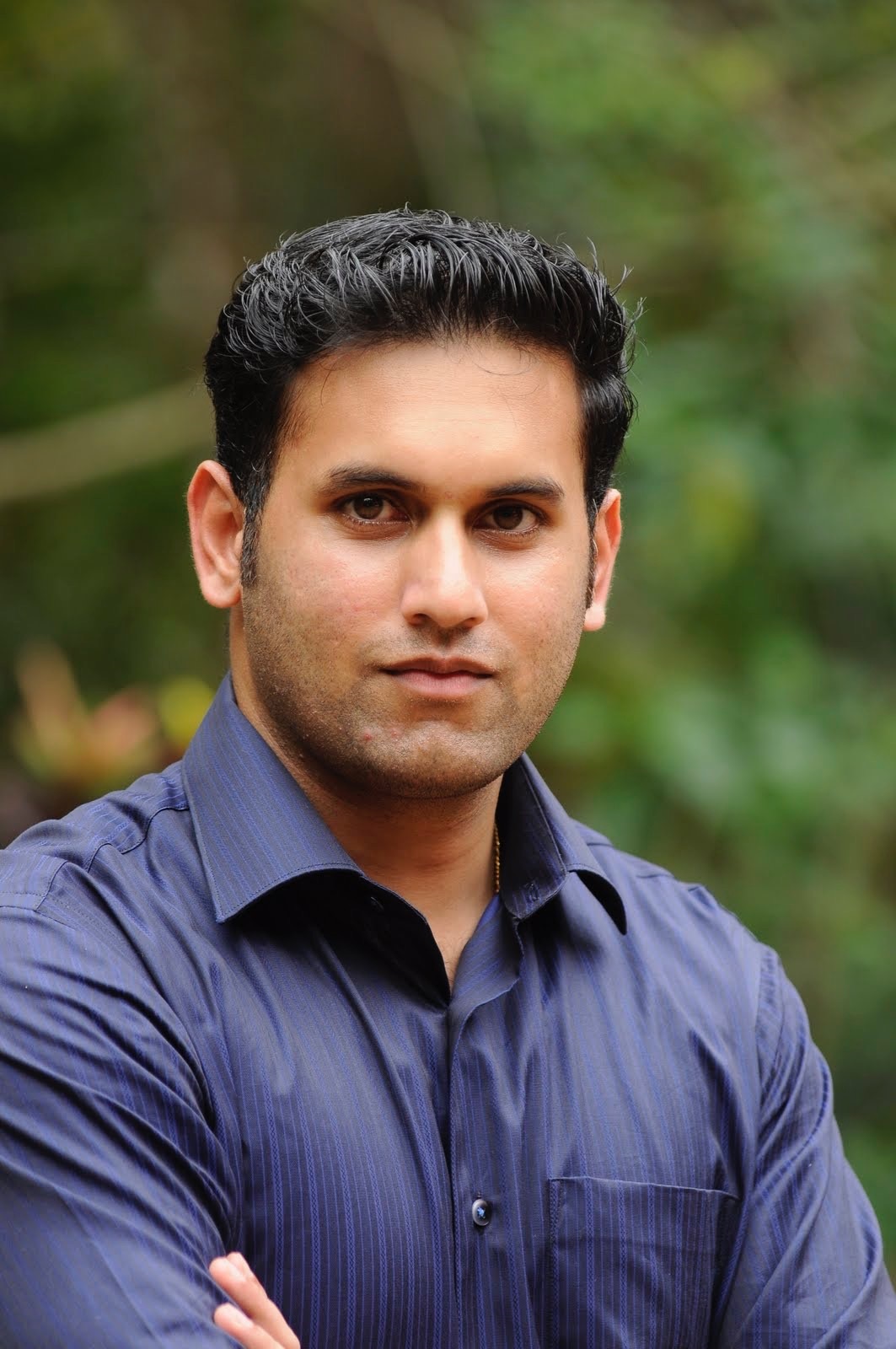}}]{Jemin George} received his M.S. (’07), and Ph.D. (’10) in Aerospace Engineering from the State University of New York at Buffalo. Prior to joining ARL in 2010, he worked at the U.S. Air Force Research Laboratory’s Space Vehicles Directorate and the National Aeronautics, and Space Administration's Langley Aerospace Research Center. From 2014-2017, he was a Visiting Scholar at the Northwestern University, Evanston, IL. His principal research interests include decentralized/distributed learning, stochastic systems, control theory, nonlinear estimation/filtering, networked sensing and information fusion. 
\end{IEEEbiography} 
\vspace{-0.5cm}
\begin{IEEEbiography}[{\includegraphics[width=1in,height=1.25in,clip,keepaspectratio]{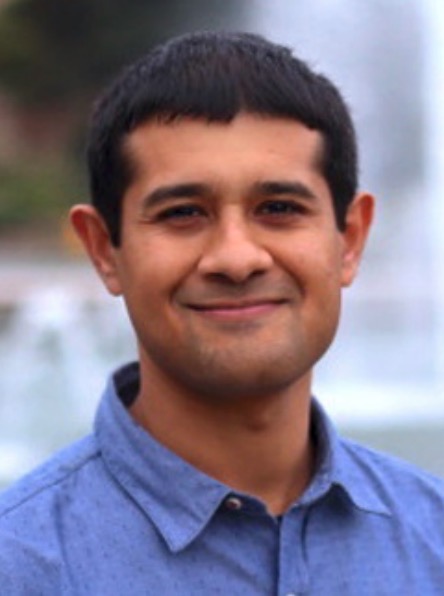}}]{Aranya Chakrabortty} received the Ph.D. degree in
Electrical Engineering from Rensselaer Polytechnic Institute, NY in 2008. From 2008 to 2009 he was a postdoctoral research associate at University of
Washington, Seattle, WA. From 2009 to 2010 he was an assistant professor at Texas Tech University, Lubbock, TX. Since 2010 he has joined the Electrical and Computer Engineering department at North Carolina State University, Raleigh, NC, where he is currently a Professor. His research interests are in all branches of control theory with applications to electric power systems. He received the NSF CAREER award in 2011.
\end{IEEEbiography} 
\end{document}